\documentclass{svjour3}                     % onecolumn (standard format)
\smartqed  % flush right qed marks, e.g. at end of proof
\usepackage{footmisc}
\usepackage{bookmark}
\usepackage{csvsimple}
\usepackage{libertine}
\usepackage[table]{xcolor}
\usepackage{booktabs,siunitx}
\usepackage{times}
\usepackage{soul}
\usepackage{mathtools}
\usepackage{url}
\usepackage[utf8]{inputenc}
\usepackage[style=base]{caption}
\usepackage{xspace}
\usepackage{amsmath}
\usepackage{amsthm}
\usepackage{fancyvrb}
\usepackage[ruled,vlined,linesnumbered]{algorithm2e}
\usepackage{algorithmic}
\usepackage{array}
\usepackage{subcaption}
\usepackage[inline]{enumitem}
\usepackage{tikz}

\usetikzlibrary{arrows,positioning,automata,shadows,fit,shapes,calc}
\usetikzlibrary{arrows.meta}
\usetikzlibrary{shapes.multipart}
\usetikzlibrary{decorations.text, shapes.misc}
\usetikzlibrary{shapes.geometric}
\newcommand{\stackk}[4]{
  \foreach \i in {1,...,#1} {
    \draw[fill=white!10] #2 ++({0.1*(#1)},{-0.1*(#1)}) ++({-0.1*\i},{0.1*\i}) rectangle +#3;
  }
  \path #2 -- +#3 node[pos=0.5] {#4};
}

\definecolor{aauBlue}{RGB}{85,158,195}
\definecolor{aauBlue2}{RGB}{67,111,129}
\usepackage{listings}% http://ctan.org/pkg/listings
\newcommand{\editcolor}{black}%{blue}%{black}

\AtEndEnvironment{example}{\qed}

\lstdefinestyle{edit}{%
basicstyle=\ttfamily\small\color{\editcolor}
}

\usepackage{graphicx}
\usepackage{template}

\newcommand{\dnot}{\ensuremath{\mathit{not}\,}}
 \journalname{Machine Learning}
\begin{document}

\title{Lifting Symmetry Breaking Constraints with Inductive Logic Programming %\thanks{Grants or other notes
%about the article that should go on the front page should be
%placed here. General acknowledgments should be placed at the end of the article.}
}
%\subtitle{Do you have a subtitle?\\ If so, write it here}

%\titlerunning{Short form of title}        % if too long for running head

\author{Alice Tarzariol         \and
        Martin Gebser  \and Konstantin~Schekotihin %etc.
}

%\authorrunning{Short form of author list} % if too long for running head

\institute{% A. 
Alice Tarzariol \at
University of Klagenfurt, Austria \\
              \email{alice.tarzariol@aau.at}             %\\
             %  if needed
           \and
%           M. 
Martin Gebser \at
           University of Klagenfurt
           and Graz University of Technology, Austria\\
           \email{martin.gebser@aau.at}  
           \and
%           K. 
Konstantin Schekotihin \at
           University of Klagenfurt, Austria\\
           \email{konstantin.schekotihin@aau.at} \and
           \emph{Correspondence to:} alice.tarzariol@aau.at 
}

\date{Received: date / Accepted: date}
% The correct dates will be entered by the editor

\maketitle

\begin{abstract}
        Efficient omission of symmetric solution candidates is essential for combinatorial problem-solving. Most of the existing approaches are instance-specific and focus on the automatic computation of Symmetry Breaking Constraints (SBCs) for each given problem instance. However, the application of such approaches to large-scale instances or advanced problem encodings might be problematic since the computed SBCs are propositional and, therefore, can neither be meaningfully interpreted nor transferred to other instances. As a result, a time-consuming recomputation of SBCs must be done before every invocation of a solver. 
        To overcome these limitations, we introduce a new model-oriented approach for Answer Set Programming that lifts the SBCs of small problem instances into a set of interpretable first-order constraints using the Inductive Logic Programming paradigm. 
        Experiments demonstrate the ability of our framework to learn general constraints from instance-specific SBCs for a collection of combinatorial problems. The obtained results indicate that our approach significantly outperforms a state-of-the-art instance-specific method as well as the direct application of a solver.
\keywords{Answer Set Programming \and Inductive Logic Programming \and Symmetry Breaking Constraints}
% \PACS{PACS code1 \and PACS code2 \and more}
% \subclass{MSC code1 \and MSC code2 \and more}
\end{abstract}
\section{Introduction}\label{sec:introduction}
% \todo{Papers must be no longer than 7 pages in total: 6 pages for the body of the paper (including all figures/tables), plus up to 1 additional page with references that do not fit within the six body pages.}
Modern declarative programming paradigms allow for relatively simple modeling of various combinatorial problems. Nevertheless, solving these problems might become infeasible when the size of input instances and, correspondingly, the number of possible solution candidates start to grow~\cite{dogalemurisc16a}. 
In many cases, these candidates are symmetric, i.e., one candidate can easily be obtained from another by renaming constants.
In order to deal with large problem instances, the ability to encode \emph{Symmetry Breaking Constraints} (SBCs) in a problem representation becomes an essential skill for programmers. 
However, the identification of symmetric solutions and the formulation of constraints that remove them is a tedious and time-consuming task. 
% Moreover, whenever an encoding must be changed, for instance, when a customer provides new requirements to a configuration problem, a programmer might need to rewrite all SBCs from scratch. 
As a result, various tools emerged that avoid the computation of symmetric solutions by, for instance, automatically finding a set of SBCs using properties of permutation groups or applying specific search methods that detect and ignore symmetric states; see \cite{margot10a,sakallah09a,walsh12a} for an overview.

Existing approaches to SBC generation can be classified into \emph{instance-specific} and \emph{model-oriented} ones. %\todo{I haven't found any name in the literature, so I invented one.} 
The former methods identify symmetries for a particular instance at hand by computing and adding ground SBCs to the problem representation \cite{DBLP:journals/constraints/CohenJJPS06,drtiwa11a,DBLP:conf/cp/Puget05}. %\todo{We can add more citations here if there is space}
Unfortunately, computational advantages do not carry forward to large-scale instances or advanced encodings, where \emph{instance-specific} symmetry breaking often requires as much time as it takes to solve the original problem.
Moreover, ground SBCs generated by \emph{instance-specific} approaches are % :
\begin{enumerate*}[label=\emph{(\roman*)}]
  \item not transferable, since the knowledge obtained is limited to a single instance;
    \item usually hard to interpret and comprehend;
    \item derived from permutation group generators, whose computation is itself a combinatorial problem; and
    \item often redundant and might result in a degradation of the solving performance.
\end{enumerate*}

In contrast, \emph{model-oriented} approaches aim to find general SBCs that depend less on a particular instance. The method presented in \cite{debobrde16a} uses local domain symmetries of a given first-order theory. SBCs are generated by identifying argument positions in atoms of a formula that comprise object variables defined over the same subset of a domain given in the input. As a result, the computation of lexicographical SBCs is very fast. However, the method considers each first-order formula separately and cannot reliably remove symmetric solutions, as it requires the analysis of several formulas at once.
%, like variants of the House Configuration Problem (HCP), e.g., things of different owners cannot be placed in one room.
The method of \cite{DBLP:conf/cpaior/MearsBWD08} computes SBCs by generating small instances of parametrized constraint programs, and then finds candidate symmetries using \textsc{Saucy} \cite{cokasama13a,saucy} -- a graph automorphism detection tool. Next, the algorithm removes all candidate symmetries that are valid only for some of the generated examples as well as those that cannot be proven to be parametrized symmetries using heuristic graph-based techniques. This approach can be seen as a simplified learning procedure that utilizes only negative examples represented by the generated SBCs.

In this work, we introduce a novel \emph{model-oriented} method for Answer Set Programming (ASP) \cite{lifschitz19a} that aims to generalize the process of discarding redundant solution candidates for instances of a target domain using Inductive Logic Programming (ILP) \cite{crdumu20a}. 
The goal is to lift the SBCs of small problem instances and to obtain a set of interpretable first-order constraints. Such constraints cut the search space while preserving the satisfiability of a problem for the considered instance distribution, which improves the solving performance, especially in the case of unsatisfiability.
The particular contributions of our work are: % paper makes the following:
\begin{itemize}
  \item We suggest several methods to generate a training set comprising positive and negative examples, allowing an ILP system to learn first-order SBCs for the problem at hand.
  \item We define the components of an ILP learning task enabling the generation of lexicographical SBCs for ASP.
  \item We analyze the features and characteristics of the results obtained by our methods, as well as the effects of language bias decisions on several combinatorial problems.% over two extensions of the pigeon-hole problem. 
  \item We present an approach that iteratively applies our method to revise constraints when new permutation group generators or more training instances become available.
  \item We conduct performance experiments on variants of the pigeon-hole problem as well as the house-configuration problem~\cite{DBLP:conf/confws/FriedrichRFHSS11}. 
  The obtained results show that the SBCs generated are easy to interpret in most of the cases, and they result in significant performance improvements over a state-of-the-art \emph{instance-specific} method as well as an ASP solver without SBCs. 
\end{itemize}

\textcolor{\editcolor}{
This work extends the previous conference paper~\cite{tagesc21a} with two additional methods to obtain the positive and negative examples for an ILP task, i.e., an alternative atom ordering criterion and a full symmetry breaking approach, along with corresponding experimental results.
% and extending the solving experiments with the constraints learned from them. 
Moreover, we provide much more detailed descriptions and analyses of % the learning 
experiments with the suggested language bias, the previous, and two new learning settings.}
% and the four different learning settings implemented.}

The structure of this paper is the following: a brief overview of the preliminaries is given in Section~\ref{sec:background}. Section~\ref{sec:approach} presents our approach, while Section~\ref{sec:implementation} describes its implementation and specifies the components of an ILP learning task. In Section~\ref{sec:learning}, we investigate observations from learning experiments conducted with our methods.
Section~\ref{sec:experiments} provides and discusses experimental results on the solving performance obtained for some combinatorial problems. Lastly, Section~\ref{sec:conclusions} concludes the paper and outlines directions for future work.

%%% Local Variables: 
%%% mode: latex
%%% TeX-master: "ijcai21"
%%% End: 

\section{Background}\label{sec:background}

This section introduces some basics and notations for ASP, symmetry breaking, and ILP.%

\subsection{Answer Set Programming}
Answer Set Programming (ASP) \cite{lifschitz19a} % \cite{gekakasc12a,gellif88b,breitr11a}
is a declarative programming paradigm based on non-monotonic reasoning and % relies on
the stable model semantics \cite{gellif91a}. 
Over the past decades, ASP has attracted considerable interest thanks to its elegant syntax, expressiveness, and efficient system implementations, showing promising results in numerous domains like, e.g., industrial, robotics, and biomedical applications \cite{ergele16a,fafrsctate18a}. 
We will briefly define the syntax and semantics of ASP, and refer the reader to \cite{gekakasc12a,lifschitz19a} for more comprehensive introductions.

\paragraph{Syntax.} 
An ASP program $P$ is a % finite 
set of \emph{rules}~$r$ of the form: 
\begin{equation*}% \label{eq:rule}
  a_0 \gets a_1, \dots, a_m, \dnot a_{m+1}, \dots, \dnot a_n
\end{equation*}
where $\mathit{not}$ stands for \emph{default negation} and $a_i$, for $0\leq i \leq n$, are atoms. 
An \emph{atom} is an expression of the form $p(\overline{t})$, where $p$ is a predicate, $\overline{t}$ is a possibly empty vector of terms, and the predicate $\bot$ (with an empty vector of terms) represents the constant \emph{false}.
Each \emph{term} $t$ in $\overline{t}$ is either a variable or a constant, and
a \emph{literal} $l$ is % either 
an atom $a_i$ (positive) or its negation $\dnot a_i$ (negative). 
%A complement of a literal $l$ is denoted as $\overline{l}$.
% Given a rule~$r$, %  of the form~\eqref{eq:rule}, 
The atom $a_0$ is the \emph{head} of a rule~$r$,
denoted by $H(r)=a_0$,
and the \emph{body} of~$r$ includes the positive or negative, respectively,
body atoms $B^+(r) = \{a_1, \dots, a_m\}$ and $B^-(r) = \{a_{m+1}, \dots, a_n\}$.
% set of atoms $B(r)=B^+(r)\cup B^-(r) = \{b_1,\dots,b_n,\} \cup \{c_{1},\dots,c_m\}$ is the \emph{body}, where $B^+(r)$ and $B^-(r)$ are disjoint sets of positive and negative body atoms, respectively.
%
A rule~$r$ is called a \emph{fact} if $B^+(r)\cup B^-(r)=\emptyset$, and a \emph{constraint} if $H(r) = \bot$.
% If any variable appearing in $H(r) \cup B^-(r)$ also appears in $B^+(r)$, then this rule is \emph{safe}. A program comprising only safe rules is also safe and, therefore, the answer sets of $P$ coincide with the answer sets of $P_{grd}$, which is a \emph{ground} program obtained by replacing all variables in $P$ with constants appearing in $P$. 

\paragraph{Semantics.}
The semantics of an ASP program $P$ is given in terms of its \emph{ground instantiation} $P_{\mathit{grd}}$,
replacing each rule $r\in\nolinebreak P$ with instances obtained by substituting the variables in~$r$ by constants occurring in~$P$.
% Let $\mathcal{A}$ be a set of ground literals appearing in $P_{grd}$.
Then, an \emph{interpretation} $\mathcal{I}$ is a set of (\emph{true}) ground atoms occurring in $P_{\mathit{grd}}$ that does not contain~$\bot$. 
% Hence, each atom $a_i \in \mathcal{I}$ is considered as true and every $a_j \notin \mathcal{I}$ is false.
An interpretation $\mathcal{I}$ \emph{satisfies} a rule $r \in P_{\mathit{grd}}$ if $B^+(r) \subseteq \mathcal{I}$ and $B^-(r) \cap \mathcal{I}=\emptyset$ imply $H(r) \in \mathcal{I}$, and
$\mathcal{I}$ is a \emph{model} of $P$ if it satisfies all rules $r\in P_{\mathit{grd}}$. 
A model $\mathcal{I}$ of~$P$ is \emph{stable} if it is a subset-minimal model of the reduct $\{H(r) \gets B^+(r) \mid r \in P_{\mathit{grd}}, B^-(r) \cap \mathcal{I} = \emptyset\}$, and we denote the set of all stable models of $P$ by $\mathit{AS}(P)$.

\subsection{Symmetry Breaking}
% Reformulated a bit to avoid usage of concepts before their introduction
Most of the modern instance-specific approaches detect symmetries of a given object by representing it as an instance of the \emph{graph automorphism problem}. This problem consists of finding all edge-preserving bijective mappings of a graph vertex set to itself. It is an attractive target of reduction since this problem can be solved efficiently for many families of graphs using methods from the \emph{group theory}; see \cite{sakallah09a} for an overview.

A \emph{group} is an abstract algebraic structure $\langle G,*\rangle$ where $G$ is a non-empty set, closed under a binary associative operator $*$, such that $G$ contains an identity element, and each element has a unique inverse. 
 If the operator $*$ is implicit, the group is represented by $G$. A \emph{subgroup}  $H$ of a group  $G$ is a group such that $H \subseteq G $.
Given a set $X=\{x_1,...,x_n\}$ % be a set 
of $n$ elements, % $x_1,...,x_n$, we can define
a \emph{permutation} of $X$ is a bijection that rearranges its elements.
The \emph{symmetric group}  $S_X$ is defined by the set of all the $n!$ possible permutations of $X$, and its subgroups are called \emph{permutation groups}.
In \emph{cycle notation}, we  represent a permutation as a product of disjoint cycles, where each cycle $\begin{pmatrix}
  x_1 & x_2 & x_3 & \dots & x_k 
\end{pmatrix}$
means that the element $x_1$ is mapped to $x_2$, $x_2$ to $x_3$, and so on, until $x_k$ is mapped back to $x_1$; the elements mapped to themselves are not contained in the cycles.

Given a group $G$ and a set $X$, the \emph{group action} $\alpha: G \xrightarrow{} S_X$ defines a permutation of the elements in $X$ for each $g \in G$.
Then, each permutation $\alpha(g)$ induces a partition on $X$, $\mathcal{P}^{\alpha(g)}(X)$, whose cells are called \emph{orbits of $X$ under $\alpha(g)$}.
The cells of the finest partition on~$X$ that refines $\mathcal{P}^{\alpha(g)}(X)$ for each $g \in G$, $\mathcal{P}^{G}(X)$, are the \emph{orbits of $X$ under group $G$}.
% To obtain the orbits of $X$ under the whole group $G$, $\mathcal{P}^{G}(X)$, we have to compute the orbits of each element in $G$ and then apply union for all of them.

\begin{example}\label{ex:1}
  Given a set $\{a, b, c, d, e\}$ of atoms ordered lexicographically and a permutation $\pi = (a \; d \; e ) \; ( b \;  c )$, let us consider a random interpretation $\{a, c\}$,
  respectively, the integer $00101$.
  To find its symmetries with respect to $\pi$, we repeatedly apply the permutation to $\{a, c\}$ until no new interpretation is obtained, e.g., $a\mapsto d$ and $c \mapsto b$ yielding $\{b,d\}$, then, $d\mapsto e$ and $b \mapsto c$ yielding $\{c,e\}$, and so on.
  Thus, we get the orbit
  $\{\{a, c\}, \{b, d\}, \{c, e\}, $ $\{a, b\}, \{c, d\}, \{b, e\}\}$, respectively,  $\{00101, 01010, 10100, 00011, 01100, 10010 \}$ with the integer representation.
\end{example}

Letting $G$ be a permutation group for a set $X$ of ground atoms,
the orbits of the set of interpretations over $X$ under $G$ identify equivalence classes of the truth assignments for~$X$.
% is and $A_X$ be the set of all the truth assignments for $X$. Then the orbits of $A_X$ under $G$ identify equivalence classes on the truth assignments of $X$: in other words, they identify classes of symmetric assignments.
% Given a permutation group $G$ together with a total order of the elements in $X$, 
When taking truth assignments as binary integers
determined by some total order of the elements in $X$,
the \emph{lex-leader} scheme consists of specifying a set of Symmetry Breaking Constraints (SBCs) that may eliminate some interpretations but keep the smallest element in terms of the associated integer for each orbit. %  of $X$ under $G$.
If the SBCs eliminate all symmetric assignments and preserve the smallest element of each orbit only, we obtain \emph{full symmetry breaking}.
However, dealing with up to $n!$ permutations of $n$ variables explicitly would be an infeasible approach, and a more efficient alternative is to focus on a subset of $G$ to refine $\mathcal{P}^{G}(X)$. If this leads to a partition finer than $\mathcal{P}^{G}(X)$, then we get \emph{partial symmetry breaking}. In this case, the SBCs preserve the smallest element as a representative of each orbit but also other symmetric interpretations.
Considering a set of \emph{irredundant generators} $K$ of $G$ is an effective heuristic for partial symmetry breaking since such generators represent $G$ compactly.
% KS: The old formulation was imprecise (see Theorem 10.3.2 in the Handbook - powers of elements in $K$)
A set $K \subset G$ of elements in a group $\langle G,*\rangle$ is a set of \emph{generators} for $G$ if every element of $G$ can be expressed as a combination of finitely many elements of $K$ under the group operation. 
%the iterated application of $*$ starting from elements of $K$ yields $G$, and
% The set of generators 
$K$ is \emph{irredundant} if no proper subset of it is a set of generators for~$G$. 
%a full SBC eliminates all but one solution, while a partial SBC may leave more than one element per orbit. 

\begin{example}\label{ex:1a}
  Let us consider the applications of the generator $\pi$ of the orbit obtained in Example \ref{ex:1}. For each interpretation in the orbit, we check whether $\pi$ maps it into a smaller interpretation according to the integer representation. Thus, the interpretations  $\{c, e\}$ and $\{b, e\}$ are eliminated since applying $\pi$  yields $\{a, b\}$ or $\{a, c\}$, respectively, while the interpretations $\{a, c\}$, $\{b, d\}$, $\{a, b\}$, and $\{c, d\}$ are preserved.
  As three symmetric interpretations are obtained in addition to the smallest interpretation of the orbit,
  the representative $\{a, b\}$ ($00011$ in the integer representation), 
  SBCs for direct applications of the generator~$\pi$ achieve partial symmetry breaking only.
\end{example}

%\subsubsection*{\textsc{SBASS}} 
Symmetry-Breaking Answer Set Solving (\textsc{sbass}) \cite{drtiwa11a} detects and eliminates syntactic symmetries in ASP by adding ground SBCs to an input ground program $P_{\mathit{grd}}$.
A symmetry of $P_{\mathit{grd}}$ is given by a permutation $\pi$ of ground atoms that keeps the program syntactically unchanged, i.e., $P_{\mathit{grd}}^\pi$ has the same rules as $P_{\mathit{grd}}$, where $P_{\mathit{grd}}^\pi$ is the set of rules obtained by applying $\pi$ to the head and body literals of rules in $P_{\mathit{grd}}$.
In the first step, \textsc{sbass} transforms $P_{\mathit{grd}}$ to a colored graph $\mathcal{G}_{P_{\mathit{grd}}}$ such that permutation groups of $\mathcal{G}_{P_{\mathit{grd}}}$ and their generators correspond one-to-one to those of $P_{\mathit{grd}}$.
In the second step, it uses \textsc{saucy} \cite{cokasama13a,saucy} to find a set of group generators for~$\mathcal{G}_{P_{\mathit{grd}}}$. 
Finally, for each found generator \textsc{sbass} constructs a set of SBCs based on the lex-leader scheme and appends them to $P_{\mathit{grd}}$.
Given the modified ground program, an ASP solver is provided means to avoid symmetric answer sets. 

\begin{example}\label{ex:2}
  To illustrate how \textsc{sbass} works, let us consider the pigeon-hole problem,
  which is about checking whether $p$ pigeons can be placed into $h$ holes
  such that each hole contains at most one pigeon.
  An encoding in ASP of this problem is:
  \begin{Verbatim}[fontsize=\verbatimsize]
   pigeon(X-1) :- pigeon(X), X > 1.
   hole(X-1) :- hole(X), X > 1.
   {p2h(P,H) : hole(H)} = 1 :- pigeon(P).
   :- p2h(P1,H), p2h(P2,H), P1 != P2.
  \end{Verbatim}
  It takes as input the ground facts \texttt{pigeon(}$p$\texttt{).} and \texttt{hole(}$h$\texttt{).} For example, solving the instance with $p=3$ and $h=3$ leads to six answer sets: 
  \begin{lstlisting}
    $\mathit{AS}_1$ = {p2h(1,1), p2h(2,2), p2h(3,3)} = 100010001
    $\mathit{AS}_2$ = {p2h(1,1), p2h(2,3), p2h(3,2)} = 010100001
    $\mathit{AS}_3$ = {p2h(1,2), p2h(2,1), p2h(3,3)} = 100001010
    $\mathit{AS}_4$ = {p2h(1,2), p2h(2,3), p2h(3,1)} = 001100010
    $\mathit{AS}_5$ = {p2h(1,3), p2h(2,1), p2h(3,2)} = 010001100
    $\mathit{AS}_6$ = {p2h(1,3), p2h(2,2), p2h(3,1)} = 001010100
  \end{lstlisting}
  The binary integer given on the right is obtained with the
  following total order of atoms:
  \begin{lstlisting}
    p2h(3,3) > p2h(3,2) > p2h(3,1) > 
    p2h(2,3) > p2h(2,2) > p2h(2,1) >
    p2h(1,3) > p2h(1,2) > p2h(1,1)
  \end{lstlisting}
  Applying \textsc{sbass} to this pigeon-hole encoding, grounded with the previous input instance, produces the following set of generators:\footnote{The generators are translated from \textsc{smodels} to  symbolic representation, as described in Section \ref{sec:implementation}.}
  \begin{lstlisting}
    $\pi_1 \; = \; \big(\,$p2h(3,2) p2h(3,3)$\,\big)$ $\big(\,$p2h(2,2) p2h(2,3)$\,\big)$ $\big(\,$p2h(1,2) p2h(1,3)$\,\big)$
    $\pi_2 \; = \;  \big(\,$p2h(3,1) p2h(3,3)$\,\big)$ $\big(\,$p2h(2,1) p2h(2,3)$\,\big)$ $\big(\,$p2h(1,1) p2h(1,3)$\,\big)$ 
    $\pi_3 \; = \; \big(\,$p2h(2,3) p2h(3,3)$\,\big)$ $\big(\,$p2h(2,2) p2h(3,2)$\,\big)$ $\big(\,$p2h(2,1) p2h(3,1)$\,\big)$ 
    $\pi_4 \; = \;  \big(\,$p2h(1,1) p2h(3,3)$\,\big)$ $\big(\,$p2h(2,1) p2h(2,3)$\,\big)$ $\big(\,$p2h(1,3) p2h(3,1)$\,\big)$
        $\quad\big(\,$p2h(1,2) p2h(3,2)$\,\big)$
\end{lstlisting}
According to the lex-leader scheme, we should discard all answer sets but $\mathit{AS}_6$, 
which is the only answer set such that applying either generator 
does not lead to a greater interpretation:% 
\begin{lstlisting}
 $\pi_1(\mathit{AS}_6) = \mathit{AS}_4 \quad  {\rightarrow} 
  \quad \mathit{AS}_6 \leq \mathit{AS}_4 \text{ \textnormal{since}}$ p2h(3,1) $=$ p2h(3,1)$\text{\textnormal{ and}}$ 
                                    p2h(2,3) $>$ p2h(2,2)
 $\pi_2(\mathit{AS}_6) = \mathit{AS}_1 \quad  {\rightarrow}  \quad \mathit{AS}_6 \leq \mathit{AS}_1 \text{ \textnormal{since}}$ p2h(3,3) $>$ p2h(3,1)
 $\pi_3(\mathit{AS}_6) = \mathit{AS}_5 \quad  {\rightarrow}  \quad \mathit{AS}_6 \leq \mathit{AS}_5 \text{ \textnormal{since}}$ p2h(3,2) $>$ p2h(3,1)
 $\pi_4(\mathit{AS}_6) = \mathit{AS}_6 \quad  {\rightarrow}  \quad \mathit{AS}_6 \leq \mathit{AS}_6 $
\end{lstlisting}
%                         $\qquad \qquad \qquad \qquad $ p2h(2,2) $>$ p2h(2,1)
Therefore, with this instance, we obtain full symmetry breaking by applying the lex-leader scheme for the irredundant generators returned by \textsc{sbass}. However, symmetric solutions can be preserved for other inputs. E.g., with $p=3$ and $h=4$, two answer sets are preserved, while the generators describe a common cell with all answer sets and a single representative.
\end{example}

\subsection{Inductive Logic Programming}
Inductive Logic Programming (ILP) \cite{crdumu20a} is a form of machine learning whose goal is to learn a logic program that explains a set of observations in the context of some pre-existing knowledge.
Since its foundation, the majority of research in the field addresses Prolog semantics \cite{metagol,muggleton95a,srinivasan01a}, even though applications in other logic paradigms appeared in the last years.  
The most expressive ILP system for ASP is \emph{Inductive Learning of Answer Set Programs} (\textsc{ilasp}) \cite{larubr14a,ilasp}, which can be used to solve a variety of ILP tasks.
% for whom several releases have been developed, extending its learning expressiveness \cite{larubr14a,larubr16a,larubr18b}

In ILP, a learning task $\langle P,E,H_M \rangle$ is defined by three elements: a background knowledge $P$, a set of (positive and negative) examples $E$, and a hypothesis space $H_M$, which defines all the rules that can be learned. The learned hypothesis is a subset of the hypothesis space that satisfies a specified learning setting: for \textsc{ilasp}, the setting is \emph{learning from answer sets} \cite{larubr14a}. 
Before defining it, we introduce the terminology used by \textsc{ilasp} authors. 
A \emph{Partial Interpretation} (PI) is a pair of sets of atoms, $e_{pi}=\langle T, F\rangle$, called \emph{inclusions} ($T$) and  \emph{exclusions} ($F$), respectively.
Given a (total) interpretation $\mathcal{I}$ and a PI $e_{pi}$, we say that $\mathcal{I}$ \textit{extends} $e_{pi}$ if $ T \subseteq \mathcal{I}$ and $F \cap \mathcal{I} = \emptyset$.
We can augment $e_{pi}$ with an ASP program $C$ to obtain a  \emph{Context Dependent Partial Interpretation} (CDPI) $\langle e_{pi}, C\rangle$.
Given a program~$P$, a CDPI $e=\langle e_{pi}, C\rangle$, and an interpretation $\mathcal{I}$,  we say that  $\mathcal{I}$ is an \emph{accepting answer set} of $e$ with respect to $P$ if $\mathcal{I} \in \mathit{AS}(P \cup C)$ such that $\mathcal{I}$ extends $e_{pi}$.

A learning task for \textsc{ilasp} is given by an ASP program~$P$ as background knowledge,  two sets of CDPIs, $E^+$ and $E^-$, as positive and negative examples, and the hypothesis space $H_M$ defined by a language bias $M$, which limits the potentially learnable rules.
The learned hypothesis $H \subseteq H_M$ must respect the following criteria: 
\begin{enumerate*}[label=\emph{(\roman*)}]
  \item for each positive example $e \in E^+$, there is some accepting answer set of $e$ with respect to $P\cup H$; and 
  \item for any negative example $e \in E^-$, there is no accepting answer set of $e$ with respect to $P\cup H$.
\end{enumerate*} 
If multiple hypotheses satisfy the conditions, the system returns one of the shortest, i.e., with the minimum number of literals \cite{larubr14a}.
In \cite{larubr18b}, the authors extend the expressiveness of \textsc{ilasp} by allowing noisy examples.
With this setting, if an example $e$ is not covered (i.e., there is an accepting answer set for $e$ if it is negative, and none, if it is positive), the corresponding weight is counted as a penalty. Therefore, the learning task becomes an optimization problem with two goals: minimize the size of $H$ and minimize the total penalties for the uncovered examples.
%In \cite{larubr16a}, the authors define an incremental strategy to scale with respect to the number of positive and negative examples $E$. It consists of using a subset $R \subseteq E$ of \emph{relevant examples}: initially it is empty, and then it grows until the examples in $R$ are sufficient to learn a hypothesis $H$ that covers all the elements in $E$. The set $R$ is constructed by interleaving the search for a hypothesis $H$ that covers the current $R$, with another search for a new example in $E \setminus R$ that is not covered by $H$. This process continues until $H$ doesn't cover all the examples, then it is returned as a solution.

Now, we will define the syntax of \textsc{ilasp} necessary for our work and refer the reader to the system's manual \cite{ilasp} for further details. 
A CDPI is expressed as follows:
\begin{align*}
  \texttt{\#type$($ID@W,\{Inc\},\{Exc\},\{C\}$)$.} 
\end{align*}
where \texttt{type} is either \texttt{pos} or \texttt{neg}, \texttt{ID} is a unique identifier for the example, \texttt{W} is a positive integer representing the example's weight  (if not defined, the weight is infinite), \texttt{Inc} and \texttt{Exc} are two sets of atoms, and \texttt{C} is an ASP program.
The language bias can be specified by \emph{mode declarations}, which define the predicates that may appear in a rule, their argument types, and their frequency. Since in our work we aim to learn constraints, we restrict the search space just to rules $r$ with $H(r) = \bot$.  
Hence, we only need to specify the mode declarations for the body of a rule, expressed by
\texttt{\#modeb$($R,P,$($E$))$} where \texttt{R} and \texttt{E} are optional and \texttt{P} is a ground atom whose arguments are 
placeholders of type \texttt{var(t)}  for some constant term \texttt{t}. 
In the learned rules, the placeholders will be replaced by variables of type~\texttt{t}.
The optional element \texttt{R} is a positive integer, called \emph{recall}, which specifies the maximum number of times that the mode declaration can be used in each rule.
Lastly, \texttt{E} is a condition that further restricts the hypothesis space.
We limit our interest to the \texttt{anti\_reflexive} option that works with predicates of arity 2. When using it, atoms of the predicate \texttt{P} should be generated with two distinguished argument values.

Choosing an appropriate language bias is still one of the major challenges for modern ILP systems. Whenever the bias does not provide enough limitations, the problem becomes intractable and \textsc{ilasp} might not be able to find useful constraints. In contrast, a too strong bias may exclude solutions from the search space, thus resulting in suboptimal SBCs \cite{cropdum20a}.

%This concept is called \textit{dependable learning} in contrast with the classical \textit{independent learning}. 
%Another interesting work is \textsc{Forgetgol} \cite{cropper19b}, a system that try to define some policy to forget useless information from the background knowledge. 

%%% Local Variables: 
%%% mode: latex
%%% TeX-master: "main"
%%% End: 

\section{Approach}\label{sec:approach}

We tackle combinatorial problems modeled in ASP such that the instances of a logic program $P$ are generated by a discrete and often stationary stochastic process. Such situations occur, e.g., in industrial settings where the encoding of a manufacturing system is fixed and production orders vary. In this case, every problem instance can be seen as an outcome of the process.
%
%More specifically, we consider an ASP encoding $P$ instantiated by any instance $d \in D$, where $D$ is a target domain.  
% Let $D$ be a set of representative instances sampled from the distribution, e.g., problem instances observed over time.
We assume that any instance
\begin{enumerate*}[label=\emph{(\roman*)}]
  \item % an instance $d \in D$
specifies the (\emph{true}) atoms of unary domain predicates $p_1,\dots,p_k$ in $P$, where $c_i$ is the number of atoms that hold for each $p_i$; and
  \item the satisfiability of the instance depends on the number of atoms for each domain predicate, but not on the values of their terms.
\end{enumerate*}
Thus, without loss of generality, we consider the terms for each $p_i$ to be consecutive integers from $1$ to $c_i$.

Our method exploits \emph{instance-specific} SBCs on a representative set of instances and utilizes them to generate examples for an ILP task. The learning method yields first-order constraints that remove symmetries in the analyzed problem instances as much as possible while preserving the instances' satisfiability.
We consider the following two learning settings: 
\begin{itemize}%{enumerate*}[label=\emph{(\roman*)}]
  \item \emph{enum} is a cautious setting that preserves all answer sets that are not filtered out by the ground SBCs; and 
  \item \emph{sat} setting aims to learn tighter constraints which, however, preserve at least one answer set for each instance.
\end{itemize}%{enumerate*}

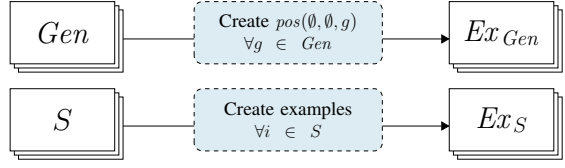
\begin{figure}[t]%\hspace*{-.7cm}
%  \begin{minipage}{.25\textwidth}
%    \centering
%\begin{tikzpicture}[>=triangle 60]
%\draw [rounded corners=0.2cm,inner sep=0pt,dashed,fill=white!20] (0,0) %rectangle ++(3,3.29) 
%node at (.3,3) [text=black!70] {\Large $\Delta$};
%%node at (2.3,2.3) [text=black!70] {Big }
%%node at (.5,-0.3) [text=black!70] {Small};
%\draw[->] (0.3,0.3) -- (2.9,3.2)node[midway,sloped,above,text=black!70]{ size};
%\draw [rounded corners=0.2cm,inner sep=0pt,fill=aauBlue!20] (0,0) rectangle ++%(1,1) 
%%node at (.2,1.2) [text=black!70] {\Large $R$}
%node at (.4,.8) [text=black!100] {$\mathit{Gen}$};
%\draw [rounded corners=0.2cm,fill=aauBlue!60] (.1,.1) rectangle ++(.5,.5) 
%node at (.1,.1) [midway,align=center] {$S$};
%\end{tikzpicture}
%\captionof{figure}{Instances domain.}
%\label{fig:instances}
%\end{minipage}
%\begin{minipage}{.25\textwidth}
  \centering
  {\LARGE
  \scalebox{0.55}{
  \begin{tikzpicture}[>=triangle 60]
%    \stackk{3}{ (-3,.8)}{(1.3,.9)}{\Large $D$}; 
%    \draw[->,dashed] (-1.5,1.1) to (0,2.2);
%    \draw[->,dashed] (-1.5,1.1) to (0,0);
  
  \draw[->] (-2.3,2.4) to (5.5,2.4);
  \stackk{3}{ (-5,1.8)}{(2.5,1.5)}{\huge $\mathit{Gen}$};
  \draw [rounded corners=0.2cm,fill=aauBlue!20,dashed] (-.6,1.8) rectangle ++(4.5,1.5) node [midway,text width=5cm,align=center] {Create $\mathit{pos}(\emptyset,\emptyset,g)$ \\ $\forall g \in \mathit{Gen}$};
  \stackk{3}{ (5.5,1.8)}{(2.5,1.5)}{\huge $\mathit{Ex}_{\mathit{Gen}}$}; 
  \draw[->] (-2.3,.3) to (5.5,.3);
  \stackk{3}{ (-5,-.3)}{(2.5,1.5)}{\huge $S$}; 
  \draw [rounded corners=0.2cm,fill=aauBlue!20,dashed] (-.6,-.3) rectangle ++(4.5,1.5) node [midway,text width=5cm,align=center] {Create examples \\ $\forall i \in S$};
  \stackk{3}{ (5.5,-.3)}{(2.5,1.5)}{\huge $\mathit{Ex}_{S}$};
  \end{tikzpicture}}}
  \captionof{figure}{ILP examples generation.}
  \label{fig:examples}
%  \end{minipage}%
\end{figure}

To compute the examples, our approach relies on small satisfiable instances (i.e., with a low value for each $c_i$), % from the set of representative instances $D$,
subdivided into two parts: $S$ and $\mathit{Gen}$. %, shown in Figure \ref{fig:instances}. 
%The former contains at least one instance that will be analyzed to detect its symmetric answer sets, and the latter contains the remaining representative instances. 
Each instance $g \in \mathit{Gen}$ defines a positive example with empty inclusions and exclusions, and $g$ as context.
These examples, denoted by $\mathit{Ex}_{\mathit{Gen}}$, guarantee that the learned constraints generalize for the target distribution since they force the constraints to preserve some solution for each $g \in \mathit{Gen}$. 
The instances $i \in S$ are used to obtain positive and negative examples, 
representing answer sets of $P \cup i$ to be preserved or filtered out, respectively,
by corresponding SBCs.
We denote their union by $\mathit{Ex}_{S}$ in Figure~\ref{fig:examples}, 
where positive examples represent whole answer sets in the \emph{enum} setting,
or like instances in $\mathit{Gen}$, consist of empty inclusions and exclusions
along with the context $i$ in \emph{sat}.

\IncMargin{1.5em}
\begin{algorithm}[t]
  \SetKwData{Left}{left}
  \SetKwData{Up}{up}
  \SetKwFunction{FindCompress}{FindCompress}
  \SetKwInOut{Input}{input}
  \SetKwInOut{Output}{output}

\Indm\Indmm
\Input{$P$, $\mathit{ABK}$, $H_M$, $\mathit{Gen}$, $S$, $m$}
\Indp\Indpp
  \BlankLine
 % $\mathit{Ex} \gets \emptyset$\;
  $\mathit{Ex}_{\mathit{Gen}} \gets \emptyset$\;
   $\mathit{Ex}_{S} \gets \emptyset$\;
   \ForEach{$g \in \mathit{Gen}$}{
%  \ForEach{$i \in S \cup \mathit{Gen}$}{
    $\mathit{Ex}_{\mathit{Gen}} \gets \mathit{Ex}_{\mathit{Gen}} \cup \{\mathit{pos}(\emptyset,\emptyset, g)\}$\;}
%   $\mathit{Ex} \gets \mathit{Ex} \cup \{\mathit{pos}(\emptyset,\emptyset, i)\}$\;}
   \ForEach{$i \in S$}{
%  $AS \gets $ Answer sets of $P \cup i \cup \mathit{ABK}$\;
  $\mathit{IG} \gets $ Set of irredundant generators for $i$\;
  \ForEach{$\mathcal{I} \in \mathit{AS}(P \cup i \cup \mathit{ABK})$}{
%    $\mathit{Ex}_{i} \gets \emptyset$\;
    $T \gets  \mathit{atoms}(\mathit{IG}) \cap \mathcal{I}$\;
    $F \gets  \mathit{atoms}(\mathit{IG}) \setminus \mathcal{I}$\;
%     $Sym \gets \mathit{lexLead}(\langle T,F\rangle, \mathit{IG})$\;
%   \uIf{Sym = True}{
  \uIf{$\mathit{lexLead}(\langle T,F\rangle, \mathit{IG})$}{
%     $\mathit{Ex}_{i} \gets \mathit{Ex}_{i} \cup  \mathit{neg}(T,F,i)$\;}
    $\mathit{Ex}_S \gets \mathit{Ex}_S \cup  \{\mathit{neg}(T,F,i)\}$\;}
     \uElseIf{m = enum}{
%    \ElseIf{m = default}{
%       $\mathit{Ex}_{i} \gets \mathit{Ex}_{i} \cup  \mathit{pos}(A,\emptyset,i)$\;
       $\mathit{Ex}_S \gets \mathit{Ex}_S \cup  \{\mathit{pos}(\mathcal{I},\emptyset,i)\}$\;
  }
  \Else{
     $\mathit{Ex}_S \gets \mathit{Ex}_S \cup \{\mathit{pos}(\emptyset,\emptyset,i)\}$\;
   }
  }
%   $\mathit{Ex}_{S} \gets \mathit{Ex}_{S} \cup \mathit{Ex}_{i}$\;
  }
%   $C \gets $ Solve ILP task $\langle P \cup \mathit{ABK}, \mathit{Ex}_{\mathit{Gen}} \cup \mathit{Ex}_{S}, S_M\rangle$\;
  $C \gets $ Solve $\langle P \cup \mathit{ABK}, \mathit{Ex}_{\mathit{Gen}} \cup \mathit{Ex}_{S}, H_M\rangle$\;
  $\mathit{ABK} \gets  \mathit{ABK} \cup C$\;
  \caption{Framework to lift SBCs with ILP}
  \label{algo}
 \end{algorithm}
 \DecMargin{1.5em}

An ILP task further requires background knowledge and a hypothesis space $H_M$. Both of them are defined by the user (for a possible instantiation, see Section \ref{subsec:ILPTask}). The \emph{background knowledge} consists of a logic program $P$ along with an \emph{Active Background Knowledge}, denoted by $\mathit{ABK}$ in Algorithm~\ref{algo}. We use $\mathit{ABK}$ to simplify the management of auxiliary predicate definitions and constraints learned so far. The \emph{hypothesis space} contains the mode declarations, and we assume it to be general enough to entail ground SBCs by learned first-order constraints.
The remaining inputs of Algorithm~\ref{algo} consist of the instances in $\mathit{Gen}$ and $S$ as well as the learning setting~$m$.
For each answer set~$\mathcal{I}$ of an instance $i\in S$ to be analyzed, the algorithm determines $T$ and $F$ by projecting $\mathcal{I}$ to the atoms occurring in $\mathit{IG}$, denoted by $\mathit{atoms}(\mathit{IG})$.
Next, in line~10, the predicate $\mathit{lexLead}(\langle T,F\rangle,$ $\mathit{IG})$ evaluates to \emph{true} if $\mathcal{I}$ is dominated, i.e., $\mathcal{I}$ can be mapped to a lexicographically smaller, symmetric answer set by means of some irredundant generator in $\mathit{IG}$.
In this case, the negative example $\mathit{neg}(T,F,i)$ is added to $\mathit{Ex}_{S}$ in order to eliminate~$\mathcal{I}$,
while $\mathit{pos}(\mathcal{I},\emptyset,i)$ or $\mathit{pos}(\emptyset,\emptyset,i)$
is taken as the positive example otherwise, depending on whether the
\emph{enum} or \emph{sat} setting is selected.
Positive examples of the form $\mathit{pos}(\emptyset,\emptyset,g)$
are also gathered in $\mathit{Ex}_{\mathit{Gen}}$ for instances $g \in \mathit{Gen}$,
and solving the ILP task at line~16 gives new constraints~$C$
to extend $\mathit{ABK}$.
% The Algorithm \ref{algo} illustrates the steps performed in our approach, requiring in input: the ASP programs $P$ and $\mathit{ABK}$, the two sets $S$ and $\mathit{Gen}$, the hypothesis space $H_M$, and the learning mode applied $m$. 
% The three arguments of functions $pos$ and $neg$ define respectively the inclusions, exclusions and context of each example; the function $\mathit{lexLead}(\langle T,F\rangle,\mathit{IG})$ checks if the truth assignment of the atoms $T \cup F$ is dominated (i.e. symmetric) according to the permutations in $\mathit{IG}$. It applies the lexicographical order of the integer terms to compare the ground atoms of $P$.

%%% Local Variables: 
%%% mode: latex
%%% TeX-master: "main"
%%% End: 

\section{Implementation}\label{sec:implementation}
\begin{figure}[t]\hspace*{-1cm}
  {\center \normalsize
\scalebox{0.8}{\begin{tikzpicture}[>=triangle 60]
  %%%  ActiveBK %%%%
\draw [draw=black]  (-4,2.5)  rectangle ++(1,1) node  [midway,text width=3.4cm,align=center] {$\mathit{ABK}$};
\draw[- ] (-3,3) -- (-2,1.5);

%%%  P %%%%
\draw [draw=black]  (-4,1)  rectangle ++(1,1) node  [midway,text width=3.4cm,align=center] {$P$};
\draw[- ] (-3,1.5) -- (-2,1.5);

%%%  I %%%%
\draw [draw=black] (-4,-.5) rectangle ++(1,1) node  [midway,text width=4cm,align=center] {$i$};
\draw[- ] (-3,0) -- (-2,1.5);
\draw[-> ] (-3.5,-.5) -- (-3.5,-3);

%%%  Gringo %%%%
\draw [rounded corners=0.2cm,fill=aauBlue!20] (-2,1) rectangle ++(2,1) node [midway,text width=3.4cm,align=center] {\textsc{Gringo}};
\draw[-> ] (0,1.5) -- (1,1.5);

%%%  P_grd %%%%
\draw [draw=black] (1,1) rectangle ++(1,1) node  [midway,text width=4cm,align=center] {$P_{\mathit{grd}}$};
\draw[- ] (2,1.5) -- (3,1.5);
\draw[-> ] (1.5,1) --  (1.5,-1);

%%%  SBASS %%%%
\draw [rounded corners=0.2cm,fill=aauBlue!20] (3,1) rectangle ++(2,1) node [midway,text width=3.4cm,align=center] {\textsc{Sbass}};
\draw[-> ] (5,1.5) -- (6,0.5);
\draw[-> ] (5,1.5) -- (6,2.5);

%%%  P_grd + SBCs %%%%
\draw [draw=black,fill=black!10]  (6,2)  rectangle ++(2.5,1) node  [midway,text width=3.4cm,align=center] {$P_{\mathit{grd}}$ + $\mathit{SBCs}$};

%%%  Permutations %%%%
\stackk{3}{ (6,0.2)}{(2.3,1)}{Permutations};
\draw[-> ] (7.25,0) -- (7.25,-3);

%%%  Clasp %%%%
\draw [rounded corners=0.2cm,fill=aauBlue!20] (0.5,-.5) rectangle ++(2,1) node [midway,text width=3.4cm,align=center] {\textsc{Clasp}};

%%%  Answer Sets %%%%
\stackk{3}{ (0.4,-2)}{(2.3,1)}{$\mathit{AS}(P_{\mathit{grd}})$}; 
\draw[- ] (2.9,-1.5) -- (6.25,-1.5);

%%%  Clingo %%%%
\draw [rounded corners=0.2cm,fill=aauBlue!20] (6.25,-2) rectangle ++(2,1) node [midway,text width=3.4cm,align=center] {\textsc{Clingo}};

%%%  Lex-lead %%%%
\draw [draw=black] (3.5,-1) rectangle ++(1.5,1) node  [midway,text width=4cm,align=center] {Lex-lead};
\draw[- ] (5,-0.5) -- (6.25,-1.5);

%%%  Assignments %%%%
\stackk{3}{ (6,-4)}{(2.3,1)}{Example data }; 
\draw[-> ] (6,-3.5) -- (-1.5,-3.5);

%%%  Ex_{i} %%%%
\stackk{3}{ (-4,-4)}{(2.3,1)}{$Ex$ of $i$}; 

\end{tikzpicture}
}}
\caption{Pipeline to compute examples from an instance $i$.}
\label{fig:pipeline}
\end{figure}
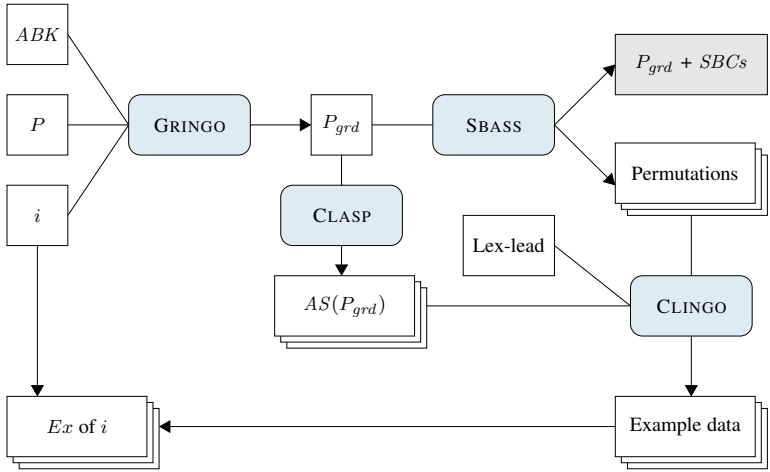

The implementation of our framework relies on \textsc{clingo}
(consisting of the grounding and solving components \textsc{gringo} and \textsc{clasp}),
\textsc{sbass} and \textsc{ilasp}, and is available at \cite{ilpsbc}. 
Figure~\ref{fig:pipeline} shows the pipeline to generate the examples for a given instance $i \in S$ (the for-loop at line 5 of Algorithm~\ref{algo}). 
First, the union of $P$, $i$, and $\mathit{ABK}$ % , represented with $P'$, 
is grounded with \textsc{gringo} to get the ground program $P_{\mathit{grd}}$ in \textsc{smodels} format.
Then, the solver \textsc{clasp} enumerates all its answer sets, obtaining $\mathit{AS}(P_{\mathit{grd}})$. Independently, \textsc{sbass} is run on  $P_{\mathit{grd}}$ with the option \texttt{--show} to output a set of irredundant permutation group generators.
This set contains the vertex permutations of $\mathcal{G}_{P_{\mathit{grd}}}$, expressed in cycle notation.
We extract the cycles defined by vertices representing atoms of $P_{\mathit{grd}}$ and transform them from \textsc{smodels} format back into their original symbolic representation (by a predicate and integer terms).
%using the integer values of terms as lexicographic ordering criterion. 
%translate them into their original representation (namely,  as predicate and terms), and order them. The translation is needed as we aim to use the integer value of the predicate terms as a sorting criterion. 
%Next, for efficiency reasons, we partition the permutations into a set $C$ of clusters according to the involved atoms. More precisely, two permutations belong to the same cluster if they share a common atom; otherwise, they are considered separately. 

Next, we identify the symmetric answer sets in $\mathit{AS}(P_{\mathit{grd}})$ by using an ASP encoding similar to the lex-leader predicate definition in \cite{sakallah09a} to evaluate SBCs.
To this end, we implement an ordering criterion to compare % allow for ordering the
atoms according to their signatures. % the integer values of their terms.
Given two ground atoms $p_1(a_1,\dots,a_n)$ and $p_2(b_1,\dots,b_m)$, the first is considered smaller than the second if: 
\begin{enumerate*}[label=\emph{(\roman*)}]
  \item $p_1$ is lexicographically smaller than $p_2$; % or
  \item $p_1=p_2$ and $n<m$; or
  \item $p_1=p_2$, $n=m$, and there are constants $a_i<b_i$ such that $a_j=b_j$ for all $0<j<i$.
% $\exists i \in [1, \dots, n] \; | \; a_i < b_i$ and $ \forall j < i \; a_j=b_j$. 
\end{enumerate*}
Our ASP encoding then checks whether an answer set $\mathcal{I}\in\mathit{AS}(P_{\mathit{grd}})$
is undominated by interpretations obtainable by applying the
symbolic representation of some irredundant generator returned by \textsc{sbass} to~$\mathcal{I}$.
% The lex-leader predicate returns the undominated atom assignments concerning the atoms in the set of generators identified by \textsc{sbass}. 

% For each $\mathcal{I} \in \mathit{AS}(P)$, we check whether an assignment according to $\mathcal{I}$ leads to unsatisfiability.
% In this case,
In case 
$\mathcal{I}$ is dominated and thus must be eliminated as a symmetric answer set,
we map it to a negative example with a unique identifier and a weight of $100$.
Due to the weights, \textsc{ilasp} returns a set of constraints even if some negative examples are not covered. Moreover, we use uniform weights so that all negative examples have the same relevance and as many as possible are to be eliminated. 
Lastly, answer sets that were not found to be dominated for any of the generators yield positive examples according to the selected setting -- \emph{enum} or \emph{sat}.
Such positive examples are unweighted so that the learned hypothesis must cover all of them.

%Moreover, since \textsc{ILASP} uses a subset of relevant examples, which grows incrementally, according to the examples' order provided,  we append the positive examples followed by the negative ones: in this way, we have a more effective search performance.

%In addition to the examples, we need to specify the background knowledge and the language bias: the former consists of $P$ and $ActiveBK$; %\footnote{We translate $P$ into an equivalent program since \textsc{ILASP} accepts a subset of ASP rules.} while the second contains a set of mode declarations.

\subsection{Alternative Atom Ordering}\label{subsec:ordering}
Let us consider sets of $n$ lexicographically ordered atoms that only differ in the values of the last terms in each atom.
For two such sets $A = \{p(\overrightarrow{x_1},a_1), \dots,$ $ p(\overrightarrow{x_n},a_n)\}$ and $B = \{p(\overrightarrow{x_1},b_1), \dots, p(\overrightarrow{x_n},b_n)\}$ of atoms,
where $\overrightarrow{x_i}$ contains all terms but the last,
the lex-leader scheme starts by checking the atoms with the greatest
$\overrightarrow{x_i}$ vectors until there are two constants $a_i \neq b_i$.
% If such index $i$ does not exists, then the two sets are equal; otherwise, the schema concludes that $A > B$ if $a_i > b_i$, or $A < B$ if $a_i < b_i$.
%
Since various configuration problems yield answer sets of this kind,
we devised an alternative atom ordering such that the lex-leader scheme starts from the smallest $\overrightarrow{x_i}$ vectors when comparing two answer sets.
To this end, an atom $p_1(a_1,\dots,a_n)$ is considered 
smaller than $p_2(b_1,\dots,b_m)$ if:
\begin{enumerate*}[label=\emph{(\roman*)}]
  \item $p_1$ is lexicographically smaller than $p_2$; % or
  \item $p_1=p_2$ and $n<m$; % or
  \item $p_1=p_2$, $n=m$, and there are constants $a_i>b_i$ such that 
        $i<n$ and $a_j=b_j$ for all $0<j<i$; or
  \item $p_1=p_2$, $n=m$, $a_i=b_i$ for all $0<i<n$, and $a_n<b_n$.
% $\exists i \in [1, \dots, n] \; | \; a_i < b_i$ and $ \forall j < i \; a_j=b_j$. 
\end{enumerate*}
    % \begin{enumerate*}[label=\emph{(\roman*)}]
    %     \item $p_1$ is lexicographically smaller than $p_2$; or
    %     \item $p_1=p_2$ and $n<m$; or
    %     \item $p_1=p_2$, $n=m$ and $\exists i \in [1, \dots, n-1] \; | \; a_i > b_i$ and $ \forall j < i, \; a_j=b_j$, or
    %     \item $p_1=p_2$, $n=m$ and $ \forall j < n, \; a_j=b_j$ and $a_n < b_n$.
    %     \end{enumerate*} 

This alternative ordering allows for more natural, undominated answer sets, as illustrated in the following example.

\begin{example}\label{ex:3}
Applying the alternative ordering criterion to the same input as described in Example~\ref{ex:2}, we get the
following total order of atoms:
\begin{lstlisting}
  p2h(1,3) > p2h(1,2) > p2h(1,1) > 
  p2h(2,3) > p2h(2,2) > p2h(2,1) > 
  p2h(3,3) > p2h(3,2) > p2h(3,1)
\end{lstlisting}
Thus, the integers associated with the answer sets become:
\begin{lstlisting}
  $\mathit{AS}_1$ = {p2h(1,1), p2h(2,2), p2h(3,3)} = 001010100
  $\mathit{AS}_2$ = {p2h(1,1), p2h(2,3), p2h(3,2)} = 001100010
  $\mathit{AS}_3$ = {p2h(1,2), p2h(2,1), p2h(3,3)} = 010001100
  $\mathit{AS}_4$ = {p2h(1,2), p2h(2,3), p2h(3,1)} = 010100001
  $\mathit{AS}_5$ = {p2h(1,3), p2h(2,1), p2h(3,2)} = 100001010
  $\mathit{AS}_6$ = {p2h(1,3), p2h(2,2), p2h(3,1)} = 100010001
\end{lstlisting}
Now the lex-leader scheme discards all but the answer set $\mathit{AS}_1$, and three permutations map $\mathit{AS}_6$ to smaller answer sets:%
\begin{lstlisting}
 $\pi_1(\mathit{AS}_6) = \mathit{AS}_4 \quad  {\rightarrow}  \quad \mathit{AS}_6 > \mathit{AS}_4 \text{ \textnormal{since}}$ p2h(1,3) $>$ p2h(1,2)
 $\pi_2(\mathit{AS}_6) = \mathit{AS}_1 \quad  {\rightarrow}  \quad \mathit{AS}_6 > \mathit{AS}_1 \text{ \textnormal{since}}$ p2h(1,3) $>$ p2h(1,1)
 $\pi_3(\mathit{AS}_6) = \mathit{AS}_5 \quad  {\rightarrow}  \quad \mathit{AS}_6 > \mathit{AS}_5 \text{ \textnormal{since}}$ p2h(1,3) $=$ p2h(1,3)$\text{\textnormal{ and}}$ 
                                    p2h(2,2) $>$ p2h(2,1)
 $\pi_4(\mathit{AS}_6) = \mathit{AS}_6 \quad  {\rightarrow}  \quad \mathit{AS}_6 \leq \mathit{AS}_6 $
\end{lstlisting}
% \begin{lstlisting}
%     $\pi_1(\mathit{AS}_6) = \mathit{AS}_4 \quad  \rightarrow  \quad \mathit{AS}_6  > \mathit{AS}_4 $ since p2h(1,3) $>$ p2h(1,2) 
%     $\pi_2(\mathit{AS}_6) = \mathit{AS}_1 \quad  \rightarrow  \quad \mathit{AS}_6 > \mathit{AS}_1 $ since p2h(1,3) $>$ p2h(1,1)
%     $\pi_1(\mathit{AS}_6) = \mathit{AS}_5 \quad  \rightarrow  \quad \mathit{AS}_6 > \mathit{AS}_5 $ since p2h(1,3) $=$ p2h(1,3) and  
%                          $\qquad \qquad \qquad \qquad $ p2h(2,2) $>$ p2h(2,1)
%     $\pi_4(\mathit{AS}_6) = \mathit{AS}_6 \quad  \rightarrow  \quad \mathit{AS}_6 \leq \mathit{AS}_6 $
% \end{lstlisting}
The answer set $\mathit{AS}_1$ contains % presents a set of 
atoms that are preserved by the general first-order constraint \texttt{:- p2h(P,H), P < H.},
which removes all other symmetric solutions.
Unlike that, when taking $\mathit{AS}_6$ as a representative solution,
we have to distinguish particular cases for the assignment of the first and the last pigeon, % from all the others,
resulting in longer and more specific constraints.  
    \end{example}

\subsection{Exploiting Generators for Full Symmetry Breaking}\label{subsec:scalable}
When investigating irredundant generators to label an answer set as a positive or negative example according to the lex-leader scheme, there can be cases where the labeling achieves partial instead of full symmetry breaking.
As illustrated in Example~\ref{ex:1a}, this is because single applications of generators yield a subset of the orbit of an interpretation only.
% This situation can happen because the lex-leader schema is sensitive to the atoms ordering and generators returned for each instance (see Example \ref{ex:1}).
Thus, we implement an alternative setting to label the examples, named \textit{fullSBCs}, which exploits generators to explore the whole orbit of symmetric interpretations for every answer set. 
For each of the obtained cells, we label the smallest answer set as a positive example and all the remaining ones as negative. 
This approach reduces the sensitivity of ILP tasks to particular irredundant generators returned by \textsc{sbass}, allowing to achieve full symmetry breaking for any instance~$i\in S$. 

%This technique takes a low computation time, because we considers just the cells identified only from answer sets.
We implement this setting by means of the \textsc{clingo} API\footnote{A complete reference documentation can be found at \url{https://potassco.org/clingo/python-api/current/}.} to interleave the solving phase, which returns a candidate answer set, with the analysis of its orbit.
Then, before continuing with the search for the next answer set, we
prohibit the explored interpretations by feeding respective constraints
to \textsc{clingo}.
% remove all the ones that were already explored, and, if the same set of examples is created from two or more distinct cells, they are written just once. 
This setting allows for reducing the number of positive examples produced, and 
as we can configure it to sample a limited subset of all answer sets, 
it is also useful for dealing with underconstrained configuration problems that yield plenty of answer sets even for very small instances.

\begin{example}\label{ex:fullsbc}
  \textcolor{\editcolor}{
    To illustrate the \textit{fullSBCs} setting, let us reconsider the pigeon-hole problem introduced in Example~\ref{ex:2},
where the instance with three pigeons and four holes leads to $24$ solutions.
Running \textsc{sbass} on this instance yields five generators, which identify a single cell since all the answer sets are symmetric. 
However, % to make this example meaningful,
we only consider the first two generators in the following,
allowing us to demonstrate the \textit{fullSBCs} approach
on an example with several, i.e., four, cells.
% , obtaining four cells.
% Note that this action is not required by our method, but we perform it only to % illustrate with a simple example how our approach works if several cells are % identified.
The generators we inspect are:}
\begin{lstlisting}[style=edit]
  $\pi_1 \; = \; \big(\,$p2h(3,2) p2h(3,3)$\,\big)$ $\big(\,$p2h(2,2) p2h(2,3)$\,\big)$ $\big(\,$p2h(1,2) p2h(1,3)$\,\big)$
  $\pi_2 \; = \;  \big(\,$p2h(3,3) p2h(3,4)$\,\big)$ $\big(\,$p2h(2,3) p2h(2,4)$\,\big)$ $\big(\,$p2h(1,3) p2h(1,4)$\,\big)$ 
\end{lstlisting}
\textcolor{\editcolor}{
Let $\mathit{AS}_1 ={}$\texttt{\string{p2h(1,3), p2h(2,2), p2h(3,4)\string}} be the first answer set found. Then, before searching for other solutions, we 
repeatedly apply $\pi_1$ and $\pi_2$ to $\mathit{AS}_1$ to obtain the whole orbit of symmetric interpretations.
% , and repeat the same action to all the different interpretations that we can obtain.
The identified answer sets are: }
\begin{lstlisting}[style=edit]
 $\qquad \qquad \qquad \, \mathit{AS}_1$ = {p2h(1,3), p2h(2,2), p2h(3,4)} = 010000101000
  $\pi_1(\mathit{AS}_1)$ =  $\mathit{AS}_2$ = {p2h(1,2), p2h(2,3), p2h(3,4)} = 001001001000
  $\pi_2(\mathit{AS}_1)$ =  $\mathit{AS}_3$ = {p2h(1,4), p2h(2,2), p2h(3,3)} = 100000100100
  $\pi_1(\mathit{AS}_2)$ =  $\mathit{AS}_1$
  $\pi_2(\mathit{AS}_2)$ =  $\mathit{AS}_4$ = {p2h(1,2), p2h(2,4), p2h(3,3)}  = 001010000100
  $\pi_1(\mathit{AS}_3)$ =  $\mathit{AS}_5$ = {p2h(1,4), p2h(2,3), p2h(3,2)} = 100001000010
  $\pi_2(\mathit{AS}_3)$ =  $\mathit{AS}_1$ 
  $\pi_1(\mathit{AS}_4)$ =  $\mathit{AS}_6$ = {p2h(1,3), p2h(2,4), p2h(3,2)}  = 010010000010
  $\pi_2(\mathit{AS}_4)$ =  $\mathit{AS}_2$ 
  $\pi_1(\mathit{AS}_5)$ =  $\mathit{AS}_3$ 
  $\pi_2(\mathit{AS}_5)$ =  $\mathit{AS}_6$  
  $\pi_1(\mathit{AS}_6)$ =  $\mathit{AS}_4$ 
  $\pi_2(\mathit{AS}_6)$ =  $\mathit{AS}_5$ 
\end{lstlisting}
\textcolor{\editcolor}{
Once we have computed all answer sets symmetric to $\mathit{AS}_1$, we produce a positive example for the smallest answer set encountered, i.e., $\mathit{AS}_2$, while the other five answer sets constitute negative examples.
Now, we can proceed with the search for the next answer set, e.g., $\mathit{AS}_7={}$\texttt{\string{p2h(1,2), p2h(2,1), p2h(3,3)\string}}, and repeat the application of generators to explore its cell, identifying another five symmetric solutions of which \texttt{\string{p2h(1,3), p2h(2,1), p2h(3,2)\string}} is the smallest. This process continues until all $24$ answer sets, partitioned into four cells with a smallest representative for each, are explored.}
\end{example}

\textcolor{\editcolor}{ 
Algorithm~\ref{algo:fullSBC} outlines the \textit{fullSBCs} approach, providing an alternative implementation of the for-loop at line 7 of Algorithm~\ref{algo}. 
%We abstract the methods used for returning a new solutions and the discarding of solutions already identified.
In the first line, we create a search control object, $\mathit{cnt}$, using the \textsc{clingo} API. This object keeps track of already identified solutions and provides the get\_new\_solution method, which returns either a new answer set $\mathcal{I}$ or \emph{false} if all solutions have been exhausted.
Similar to the previously presented approaches to example generation, we project the atoms of $\mathcal{I}$ to $\mathit{atoms}(\mathit{IG})$.
The resulting interpretation $\mathit{min}$ represents the smallest solution encountered so far in the current cell, and the set $\mathit{seen}$ keeps track of already discovered interpretations belonging to the current cell.
Starting with $\mathit{min}$, the queue $Q$ collects the interpretations to which all irredundant generators will be applied to yield new symmetric interpretations.
The while-loop at line 7 checks whether there is an interpretation, $T$, left to pop. Then, if $T$ is greater than $\mathit{min}$ (according to the applied atom ordering criterion), it constitutes a negative example, while a smaller $T$ is taken as new smallest interpretation and the previous $\mathit{min}$ instead becomes a negative example.
Only after the cell has been completely explored, the interpretation $\mathit{min}$ is eventually labeled as a positive example. Lastly, before querying $\mathit{cnt}$ for the next solution, we eliminate answer sets
subsumed by the explored interpretations in $\mathit{seen}$
from the search space of $\mathit{cnt}$.}
% The implementation of Algorithm \ref{algo:fullSBC} as well as Algorithm \ref{algo} are available at \cite{ilpsbc}.}

\IncMargin{1.5em}
\begin{algorithm}[t]
  \SetKwData{Left}{left}
  \SetKwData{Up}{up}
  \SetKwFunction{FindCompress}{FindCompress}
  \SetKwInOut{Input}{input}
  \SetKwInOut{Output}{output}
  \color{\editcolor}
\Indm\Indmm
\Input{$P$, $i$, $\mathit{ABK}$, $\mathit{IG}$}
\Indp\Indpp
  \BlankLine
  $\mathit{cnt} \gets \text{\textsc{clingo}}.\text{init}(P \cup i \cup \mathit{ABK})$\;
  \While{$\mathcal{I} \gets \mathit{cnt}.\text{\textup{get\_new\_solution}}()$ }{
  $\mathit{min} \gets \mathit{atoms}(\mathit{IG}) \cap \mathcal{I}$\;
  $\mathit{Q} \gets \text{new\_queue}()$\;
  $\mathit{Q}.\text{push}(\mathit{min})$\;
  $\mathit{seen} \gets \{\mathit{min}\}$\;
   \While{$T \gets \mathit{Q}.\text{\textup{pop}}()$} % $\mathit{Q}$ \textup{is non-empty}}
   {
%    $T \gets \mathit{Q}.\text{pop}()$\;
    \uIf{$T > \mathit{min}$}{
      $\mathit{Ex}_S \gets \mathit{Ex}_S \cup \{\mathit{neg}(T, \mathit{atoms}(\mathit{IG})\setminus T, i)\}$\;
        }
          
    \ElseIf{$T < \mathit{min}$}{
        $\mathit{Ex}_S \gets \mathit{Ex}_S \cup \{\mathit{neg}(\mathit{min}, \mathit{atoms}(\mathit{IG})\setminus \mathit{min}, i)\}$\;
        $\mathit{min} \gets T$\;}
    \ForAll{$\pi \in \mathit{IG}$}{
      \If{$\pi(T) \notin \mathit{seen}$}{
        $\mathit{Q}.\text{push}(\pi(T))$\;
        $\mathit{seen} \gets \mathit{seen} \cup \{\pi(T)\}$\;
        }
    }}
    $\mathit{Ex}_S \gets \mathit{Ex}_S \cup \{\mathit{pos}(\mathit{min}, \mathit{atoms}(\mathit{IG})\setminus \mathit{min}, i)\}$\;
    $\mathit{cnt}.\text{ignore\_solutions}(\mathit{seen})$\;
    }
  \caption{\textit{fullSBCs} method to generate $\mathit{Ex}_S$ for an instance $i$}
  \label{algo:fullSBC}
 \end{algorithm}
 \DecMargin{1.5em}

\subsection{ILP Learning Task}\label{subsec:ILPTask}
After considering the example generation, we specify components of the ILP learning task suitable for the learning of constraints.
The idea is to encode the predicates used by lex-leader symmetry breaking to order atoms and extract the maximal values for domain predicates.
Since the mode declarations of \textsc{ilasp} (v4.0.0) do not support arithmetic built-ins such as \texttt{<}, we provide auxiliary predicates in $\mathit{ABK}$ to simulate them.
  %The goal to reduce the hypothesis space to minimal information is common in ILP (cf.\ Blumer bound in reference {[Cropper \emph{et al.}, 2020]})
% We defined $\mathit{Gen}$ with trivial but representative instances. When we evaluated the constraints, if they removed all the solutions for a satisfiable instance $i \notin \mathit{Gen}$, we backtracked; namely, we forgot the rules learned so far, added $i$ in $\mathit{Gen}$, and repeated the learning.
Presupposing the presence of unary domain predicates $p_1,\dots,p_k$
with integers from $1$ to $c_i$ for each $p_i$,
$\mathit{ABK}$ defines the auxiliary predicates \texttt{max$p_i(c_i)$} for each $p_i$ and \texttt{lessThan$(t_1$,$t_2)$} for each pair of integers $1\leq t_1<t_2\leq \max\{c_i \mid i=1,\dots,k\}$.
These two predicates,
based exclusively on syntactic properties of a considered problem, are minimal for overcoming limitations of \textsc{ilasp} to learn lex-leader SBCs.
The selection of small yet representative instances for $S$ and $\mathit{Gen}$ depends on their hardness for learning.
Regarding~$S$, we pursued the strategy to empirically determine instances for which \textsc{sbass} yields a manageable number of permutation group generators. As mentioned in Section~\ref{subsec:scalable}, the irredundant generators alone sometimes achieve partial symmetry breaking,
and we selected only instances without any or a small amount of ``misclassified" answer sets.
The instances in $\mathit{Gen}$ are usually larger yet still solvable in a short running time to check that the learned constraints generalize.

The language bias of our learning task includes the mode declarations  \texttt{\#modeb$($2, $p_i($var$(t_i)))$} and \texttt{\#modeb$($1,max$p_i($var$(t_i)))$} for each domain predicate $p_i$, in which \texttt{var}$(t_i)$ is a placeholder indicating the domain for which each $p_i$ holds. 
Moreover,  for each (non-auxiliary) predicate \texttt{P} appearing in some of the generators computed for instances in $S$, we use \texttt{\#modeb$($2,P$)$},
where the domains of variables in atoms of~\texttt{P} are provided % indicated 
by a vector of the placeholders in $\{\text{\texttt{var}}(t_i) \mid i =\nolinebreak 1,\dots, k\}$,
depending on the role of \texttt{P} in the given program $P$.
In addition, we include mode declarations \texttt{\#modeb$($2,lessThan$($var$(t_i)$,\linebreak[1]var$(t_j)))$} for all  $i,j = 1, \dots, k$, with the option \texttt{anti\_reflexive} in case $i=j$.

We decided to distinguish the variables' types in the mode declarations in order to restrict the hypothesis space to rules such that a variable \texttt{X} of type~$t$ is included as an argument only in predicates defined over the same type~$t$.
To illustrate how this decision influences the search space of an ILP task, let us consider two extensions of the pigeon-hole problem introduced in Example~\ref{ex:2}, adding color and owner assignments. The pigeon-hole problem with colors associates a color with each pigeon and requires pigeons placed into neighboring holes to be of the same color.
The version with colors and owners additionally assigns an owner to each pigeon and imposes the same constraint as with the colors for owners as well.
For the pigeon-hole problem with colors, by using typed variables in the mode declarations, \textsc{ilasp} generates a search space of $1837$ rules,\footnote{For all our experiments, we used the default value (3 literals) for the \textsc{ilasp} parameter that defines the maximum number of literals that can occur together in the body of each rule of the hypothesis.} while $9169$ rules are obtained without distinguishing variables' types.
Regarding the extension to owners, this difference is even larger: $2895$ rules using typed variables versus $21406$ rules without distinguishing variables' types. 

To compare the learning performance of \textsc{ilasp}, we conducted several experiments on the pigeon-hole problem with colors and owners for a pool of instances\footnote{The collected data can be found at \cite{ilpsbc}. The experiments were run on an Intel\textsuperscript{\textregistered} i7-3930K machine under Linux (Debian GNU/Linux 10), where each run of \textsc{ilasp4} was limited to 3600 seconds.\label{foo:data}} and observed that applying our approach with typed variables in the mode declarations allows for learning constraints quicker than without distinguishing the types. When using the iterative approach described in Section~\ref{subsec:iterative}, \textsc{ilasp} took on average less than two minutes to learn the shortest constraints related to holes, colors, and owners, and always finished in less than ten minutes. In opposite, a similar ILP task defined without distinction of variable types took on average thirty minutes, with cases where no hypothesis was found within an hour.

Reducing the hypothesis space has the potential drawback of learning less efficient rules since there can be situations where stronger constraints with fewer variables are excluded.
For instance, a constraint like \texttt{:- pigeon(X), not p2h(X,X).} cannot be learned in the current setting, as the variable \texttt{X} is taken for a pigeon and a hole at the same time. 
However, we decided to use the restricted search space for our experiments in Section~\ref{sec:experiments} because it leads to much better scalability of learning and constraints that still improve the solving performance.
In fact,
the ability to learn constraints in acceptable time is important for
handling application scenarios better than with instance-specific symmetry breaking methods.

%\begin{itemize}
%    \item[+] restricts the search space, thus faster search.
%    \item[-] maybe less efficient constraints if more variables 
%\end{itemize}

\begin{example}\label{ex:4}
  To illustrate a feasible outcome of our ILP framework, % and components of the ILP task in practice,
  % we provide an example of constraints learnable 
  let us inspect the constraints learned for the pigeon-hole problem and its instance with three pigeons and three holes, as also considered in Example~\ref{ex:2}. 
  Applying the generators returned by \textsc{sbass} to the six answer sets gives one positive and five negative examples, and the resulting ILP task is as follows:
  \begin{Verbatim}[fontsize=\verbatimsize]
 %% Input encoding adapted for ILASP 
 pigeon(X-1) :- pigeon(X), X > 1.
 hole(X-1) :- hole(X), X > 1.
 0 {p2h(P,H)} 1 :- pigeon(P), hole(H).
 :- p2h(P1,H), p2h(P2,H), P1 < P2.
 :- p2h(P,H1), p2h(P,H2), H1 < H2.
 :- pigeon(P), not p2h(P,_).
  
 %% Active Background Knowledge
 lessThan(X,Y) :- pigeon(X), pigeon(Y), X < Y.
 lessThan(X,Y) :- hole(X), hole(Y), X < Y.
 maxpigeon(X) :- pigeon(X), not pigeon(X+1).
 maxhole(X) :- hole(X), not hole(X+1).

 %% Negative examples
 #neg(id1@100, {p2h(2,3), p2h(1,2), p2h(3,1)}, 
  {p2h(2,1), p2h(1,1), p2h(3,3), p2h(1,3), p2h(3,2), p2h(2,2)},
  {pigeon(3). hole(3).}).
 #neg(id3@100, {p2h(2,1), p2h(3,2), p2h(1,3)},
  {p2h(1,1), p2h(3,3), p2h(3,1), p2h(2,2), p2h(2,3), p2h(1,2)},
  {pigeon(3). hole(3).}).
 #neg(id4@100, {p2h(2,3), p2h(1,1), p2h(3,2)},
  {p2h(2,1), p2h(3,3), p2h(3,1), p2h(1,3), p2h(2,2), p2h(1,2)},
  {pigeon(3). hole(3).}).
 #neg(id5@100, {p2h(2,1), p2h(3,3), p2h(1,2)},
  {p2h(1,1), p2h(3,1), p2h(1,3), p2h(3,2), p2h(2,3), p2h(2,2)}, 
  {pigeon(3). hole(3).}).
 #neg(id6@100, {p2h(1,1), p2h(3,3), p2h(2,2)},
  {p2h(2,1), p2h(3,1), p2h(1,3), p2h(3,2), p2h(2,3), p2h(1,2)},
  {pigeon(3). hole(3).}).
   
 %% Positive example produced with enum setting
 #pos(id2, {p2h(3,1), p2h(2,2), p2h(1,3)}, {}, 
  {pigeon(3). hole(3).}).
 
 %% Language bias
 #modeb(2,p2h(var(pigeon),var(hole))).
 #modeb(2,pigeon(var(pigeon))).
 #modeb(2,hole(var(hole))).
 #modeb(1,maxhole(var(hole))).
 #modeb(1,maxpigeon(var(pigeon))).
 #modeb(2,lessThan(var(hole),var(hole)),(anti_reflexive)).
 #modeb(2,lessThan(var(pigeon),var(pigeon)),(anti_reflexive)).
 #modeb(2,lessThan(var(hole),var(pigeon))).
 #modeb(2,lessThan(var(pigeon),var(hole))).
  \end{Verbatim}
  Let us notice that the ASP input encoding in Example~\ref{ex:2} has been adapted into an equivalent one above. Such a modification is necessary because the current version of \textsc{ilasp} does not support rules like
  \texttt{\{p2h(P,H) : hole(H)\} = 1 :- pigeon(P).}
  with the conditional operator ``\texttt{:}" in the head.
  After running \textsc{ilasp}, the learned first-order constraints are:
  \begin{Verbatim}[fontsize=\verbatimsize]
 :- p2h(X,Y), lessThan(Z,Y), maxpigeon(X).
  % do not assign the pigeon with the max label to a hole
  % other than the first one
 :- p2h(X,Y), lessThan(X,Y), lessThan(Y,Z).
  % for all but the last hole, do not assign a pigeon with 
  % a smaller label to the hole
   \end{Verbatim}
  \end{example}

\subsection{Iterative Learning}\label{subsec:iterative}
Inspired by the \textit{lifelong learning} approach \cite{crdumu20a}, we apply our framework incrementally to a split learning task. 
This idea is especially useful if the ASP encoding presents several symmetries, where some of them are independent of the others. 
The iterative approach simplifies the learning task by exploiting the incremental applicability of ILP: first, it solves a subtask to identify a subset of symmetries, and before addressing the remaining ones, we integrate the constraints just learned into the background knowledge.
To this end, we % partition the instances and 
divide the hypothesis space for programs with three or more types of variables in the language bias. %in the first ILP run, the mode declarations are restricted to two types of variables, say $t_1$ and~$t_2$, and then they are progressively extended to further types from $t_3$ to~$t_k$.
%More specifically, 
Then,
in the first step, we provide a set~$S$ of instances to address their symmetries involving only two types of variables and define the search space with mode declarations restricted to the two types of variables considered.
Next, we solve the ILP subtask and append the learned constraints to $\mathit{ABK}$. In the following steps, we repeat the procedure and analyze the same or different instances in $S$ for symmetries going beyond those already handled by solving ILP subtasks with the mode declarations progressively extended to further types of variables.

%For example, for the pigeon-hole problem with color and owner assignemnts, we can easily identify the symmetries related exclusively to the pigeon's placements by considering the instances with one color.
To illustrate a concrete application scenario, reconsider the pigeon-hole problem with color and owner assignments, introduced in Section~\ref{subsec:ILPTask}. For this problem, the search space is split into three incremental parts: 
\begin{itemize}
   \item the first is limited to predicates whose atoms exclusively include variables of the types \texttt{pigeon} and \texttt{hole}, 
   \item the second part extends mode declarations by allowing atoms with variables of the type \texttt{color} too, and, finally,
   \item the third step includes variables of the type \texttt{owners}.
\end{itemize}
Initially, $S$ contains instances with only one color and owner so that our framework produces examples entailing symmetries related exclusively to the pigeons' placement.  Next, we append the learned constraints to $\mathit{ABK}$ and repeat the procedure by redefining $S$ with instances with one owner but more than one color. 
Lastly, we analyze instances in $S$ without restrictions on the numbers of colors and owners while considering the whole language bias. 
In this way, \textsc{ilasp} can learn new symmetries using predicates that involve more types of variables, as the language bias is progressively extended until it reaches the whole set of mode declarations.

By applying the iterative approach, the learning task can be decomposed into smaller and easier ILP subtasks.
For an indication of the practical impact on the size of the search space(s), we note that 
\textsc{ilasp} generates $1040$ rules for 
variables of the types \texttt{pigeon} and \texttt{hole} only,
$1837$ rules when variables of the type \texttt{color} are added,
and $2895$ rules with the full language bias for the pigeon-hole problem with colors and owners.
That is, the search space for the ILP subtask in the last iteration includes the same rules as generated when addressing the full language bias in a single step, yet the background knowledge may already be extended by constraints reducing the number of (negative) examples still to investigate.
% the pigeon-hole problem with colors and owners: if we immediately consider the whole language-bias, we get a search space of $2895$ rules. 
% In contrast to that, the mode declarations that involve only variables of the types \texttt{pigeon} and \texttt{hole} yield $1040$ rules, while extending also the type \texttt{color}, it produces $1837$ rules.
We assessed the impact of the iterative approach in several experiments% for a pool of instances
\footref{foo:data} % \footref{foo:settings}
and observed that it allows us to learn constraints much quicker than when tackling all symmetries in a single pass. By splitting the learning task, \textsc{ilasp} took on average less than two minutes to learn constraints related to pigeons and holes, then colors, and finally owner assignments.
Unlike that, the ILP task that addresses the full language bias directly took on average more than thirty minutes to return the shortest hypothesis, where in some cases the search was not finished within one hour.

Splitting the learning task has the potential drawback that some of the symmetries can be lost in the process, as the updated $\mathit{ABK}$ is considered in subsequent calls to \textsc{sbass} for identifying remaining symmetries. However, for the combinatorial problems investigated in our experiments in Section~\ref{sec:experiments}, the results showed that even in case we learn constraints handling a subset of all problem symmetries, the solving performance benefits substantially.

%\begin{itemize}
   %\item[+] exploit ILP nature of reuse knowledge. 
   %\item[+] especially useful for independent symmetries.
   %\item[+] decompose the problem, thus faster search.
   %\item[-] we may lose some symmetries in the process
%\end{itemize}

% with the other variables.
%This iterative approach can speed up the learning procedure and required less than a minute of running time for each iteration on the combinatorial problems investigated in Section~\ref{sec:experiments}.

%%% Local Variables: 
%%% mode: latex
%%% TeX-master: "main"
%%% End: 

\section{Learning Performance}\label{sec:learning}

We tested the different settings of our implementation over the two extensions of the pigeon-hole problem described in Section~\ref{subsec:ILPTask}.  For every setting, we used the same initial set of instances in $\mathit{Gen}$, auxiliary constraints in $\mathit{ABK}$, and mode declarations in the language bias (split to apply the iterative approach). 
For keeping the number of instances in $\mathit{Gen}$ moderate,
we hand-picked a few (satisfiable) instances to start from,
applied our iterative learning approach,
and then validated the learned constraints on other satisfiable instances as well.
% We defined $\mathit{Gen}$ in a pre-processing step: we started with an initial guess, applied our approach to learn constraints and then 
%  validate them over a set of satisfiable instances.
The instances for which learned constraints led to unsatisfiability were then also added to $\mathit{Gen}$, and we repeated the learning phase until all instances were found satisfiable together with the learned first-order SBCs. %final hypothesis of first-order constraints. 

In the following, we report representative results and conclusions drawn from the instances and records of learning experiments provided in our repository \cite{ilpsbc}.

%\todo[inline]{introduce this section: what we want to show here. We get constraints that help the search. They are not always SBC. We tested the pigeon owner problem.}

\subsection{Enum vs Sat Setting}\label{subsec:sat}
The difference between the \emph{enum} and \emph{sat} setting lies in the positive examples generated for the instances in $S$: in the first setting, we explicitly list all undominated answer sets as positive examples, while the second produces just a general positive example with empty inclusions and exclusions.
That is, the \emph{sat} setting abstracts over undominated answer sets, as they are neither labeled as positive nor negative examples in the ILP task.
In this case, \textsc{ilasp} aims at eliminating as many symmetric answer sets as possible while preserving the satisfiability of a given instance.
This even means that the preserved answer sets, required in view of the general positive example, might belong to negative examples but are not covered by the learned constraints.
In this way, we may, in general, learn alternative constraints that preserve some specific pattern of solutions appearing in all satisfiable instances, regardless of symmetries, while representative solutions can be lost. 

For example, in the first ILP iteration on the pigeon-hole problem with colors and owners,
the instance with three pigeons and four holes (and only one color and owner) gives $24$ answer sets,
$22$ of which are labeled as negative.
In the \emph{enum} setting, \textsc{ilasp} finds optimal constraints removing $12$ negative examples and thus returns a hypothesis that applied to the same instance leaves $12$ answer sets.
In contrast, the \emph{sat} setting enables learning of stronger constraints by \textsc{ilasp},
which preserve $2$ answer sets only, both labeled as negative examples. 

The complexity of the ILP task depends on the possibility of covering all negative examples. Since with \emph{enum} we have tighter conditions on the candidate hypotheses, the search space is smaller than in the \emph{sat} setting. Hence, % since we define all the negative examples with weights,
the optimization problem regarding (weighted) negative examples addressed by \textsc{ilasp} takes in general longer for \emph{sat}, but only if many negative examples cannot be covered even under relaxed conditions on the candidate hypotheses.
On the other hand, if the language bias permits many hypotheses covering all or most of the negative examples, an ILP task is usually quickly solved with the \emph{sat} setting.
E.g., the instance with three pigeons, four holes, one color and owner took $72.8$ seconds to be solved in the \emph{enum} (eliminating $12$ out of $22$  symmetric answer sets) and just $27.7$ seconds in the \emph{sat} setting (eliminating $20$ symmetric answer sets and the $2$ unlabeled ones).

%\begin{itemize}
    %\item[+] It can focus on a specific patter of solutions that appear always. 
    %\item[+] abstract some answer sets: the undominated one are unlabeled, thus not considered in the search. Just remove as many AS as possible, while preserving the satisfiability of the instances.
    %\item[+] The preserved answer set may be negative. 
    %\item[-] larger search space since there are more complete hypothesis (because of general pos example), while the consistent hypothesis are an optimazion problem. So if many uncoverd as, it takes longer.
    %\item[-] It can remove many solutions    
%\end{itemize}
\subsection{Alternative Atom Ordering}\label{subsec:ordExperim}
When answer sets for the combinatorial problem analyzed have the property described in Section~\ref{subsec:ordering}, our alternative ordering criterion for the lex-leader scheme may distinguish the representative and symmetric solutions. Hence, ILP tasks can be solved with shorter constraints than for the default atom ordering. 
For instance, the setting illustrated in Example~\ref{ex:4} yields the representative
answer set \texttt{\{p2h(1,1),p2h(2,2),p2h(3,3)\}} instead of \texttt{\{p2h(1,3),p2h(2,2),p2h(3,1)\}}. 
This allows \textsc{ilasp} to learn the short constraint \texttt{:- p2h(X,Y), lessThan(Y,X).},
expressing that no pigeon can be placed into a hole smaller than its label,
which leaves just one answer set for instances with an equal number $p=h$ of pigeons and holes.

Given that the positive examples kept after checking direct applications of the irredundant generators returned by \textsc{sbass} heavily depend on the computed generators, we found that often more positive examples are produced than for the default atom ordering.
Namely, the generators preserve more symmetric solutions than the default ordering for the extensions of the pigeon-hole problem to colors as well as colors and owners.
This leads to weaker (although shorter and easier to interpret) constraints, and better-suited ways of aligning generators with symbolic atom representations would be of interest.
%\begin{itemize}
%    \item[+] If the setting is the one described in the previous section (set of atoms with same cardinality and args), the ordering of AS is conduct as expected (i.e. minimize the smaller atom first).
 %   \item[+] in some cases we obtain shorter constraints 
  %  \item[-] since it heavily depends on the generators returned by \textsc{sbass}, more positive examples are produced in general.  
%\end{itemize}
\subsection{Exploiting Generators for Full Symmetry Breaking}
Section~\ref{subsec:scalable} describes an alternative implementation for labeling answer sets as positive or negative examples, called the \textit{fullSBCs} setting. 
We can see the effects of always labeling the answer sets according to full SBCs on the same scenario as discussed in Section~\ref{subsec:sat}: 
%in the first iteration of the pigeon-hole problem with colors and owners, the instance with three pigeons and four holes (and only one color and owner) 
%produces $24$ answer sets, among which $2$ were labeled as positive.
instead of $22$ negative and $2$ positive examples generated with the \emph{enum} setting, \textit{fullSBCs} returns just one positive example, i.e., the representative of the single cell characterized by the generators of \textsc{sbass}. As a consequence, 
instead of $72.8$ seconds to return a hypothesis that produces $12$ of the original $24$ answer sets, with the ILP task defined based on \textit{fullSBCs}, \textsc{ilasp} took $21.4$ seconds to find a hypothesis that preserves only $4$ answer sets.

We note that reducing the number of examples for an ILP task generated by some of our settings has a limited impact on \textsc{ilasp}, as its latest versions implement mechanisms to scale with respect to the number of examples \cite{larubr16a,law21a}. However, for instances with many answer sets, the \textit{fullSBCs} approach can be helpful because equivalent answer sets need not be exhaustively computed by \textsc{clingo}.

%In the previous example, the generators produced just one cell of symmetric answer sets. Now, let us consider another situation of the same problem: from an $\mathit{ABK}$ with the constraints:
%\begin{Verbatim}[fontsize=\verbatimsize]
%  :- assign(V2,V3), lessthan(V1,V2), maxhole(V3).
%  :- assign(V3,V2), lessthan(V1,V2), maxpigeon(V3).
%  :- assignColor(V3,V1), lessthan(V1,V3), lessthan(V2,V1).
%\end{Verbatim}
%the instance with two owners, two colors, three pigeons and five holes produces $30$ answer sets, among which half of them are labeled as positive from the \emph{enum} approach. Using \textit{fullSBCs}, only three negative and three positive examples are produced, because the same inclusions and exclusions would have been generated from the other four cells. We point out that \textsc{ilasp} is already able to scale with respect to the number of examples \cite{larubr16a}. This mechanism iteratively sample the examples until all of them are covered 

%\begin{itemize}
 %   \item[+] Produce less examples
  %  \item[+] Less variability in the learned constraints since is less sensitive to partial SBCs. 
   % \item[-] Sometimes you lose the possibility to apply the iterative mode because the first steps are very strict. 
%\end{itemize}

%%% Local Variables: 
%%% mode: latex
%%% TeX-master: "main"
%%% End: 

\section{Solving Experiments}\label{sec:experiments}
To evaluate our approach and the implementation design, we applied it to a series of combinatorial search problems.
For each considered problem, we compared the running time of the original encoding, the version extended with our learned constraints, and the \emph{instance-specific} approach of \textsc{sbass}.
The learned constraints depend on the instances used in $S$ and $\mathit{Gen}$ as well as how we apply the iterative learning approach.  In the following, we report results for the constraints with good performance learned applying the definitions of Section~\ref{subsec:ILPTask}.\footnote{Detailed settings are provided at \cite{ilpsbc}.\label{foo:settings}}
We ran our tests on an Intel\textsuperscript{\textregistered} i7-3930K machine under Linux (Debian GNU/Linux 10), where each run was limited to 900 seconds and 20 GB of memory.

In Table~\ref{table:ph} to Table~\ref{table:hc}, the satisfiable instances are shown in grey rows, while the white rows contain unsatisfiable instances. The column \textsc{base} refers to \textsc{clingo} (v5.5.0)
run on the original encoding, while \textsc{Enum}, \textsc{Sat}, \textsc{Ord}, and \textsc{Full} report results for the original encoding augmented with first-order constraints learned in the \emph{enum}, \emph{sat}, and \emph{enum} with alternative atom ordering or \textit{fullSBCs}  setting, respectively.
The time required by \textsc{sbass} to compute ground SBCs is given in the corresponding column, and \textsc{clasp}$^\pi$ provides the solving time obtained with these ground SBCs.
% consists of the time in the homonymous column and in \textsc{clasp}$^\pi$: the former contains  the grounding and pre-processing time needed to compute the ground SBCs, while the latter contains the time required to solve $P_{grd}$ and the ground constraints.
Runs that did not finish within the time limit of 900 seconds are indicated by TO entries.

We first tested the pigeon-hole problem, working without any division and iterative analysis of the language bias: the four learning settings led to similar performance constraints, although the ones obtained with the alternative ordering were shorter, as mentioned in Section~\ref{subsec:ordExperim}.
The running time comparison in Table~\ref{table:ph} shows that all the settings of our approach bring about a similar speedup for solving satisfiable as well as unsatisfiable instances.
In fact, the vast problem symmetries are cut by the learned first-order constraints. This is particularly important in case of unsatisfiability,
where runs on the original encoding without additional constraints do not finish within the time limit.
% The original encoding identifies satisfiable instances with a considerable number of pigeons and holes, but it fails to detect not trivial unsatisfiable instances.
While \textsc{sbass} also manages to handle the two smallest instances,
the computation of permutation group generators becomes too expensive when the
instance size grows,
in which case we cannot run \textsc{clasp}$^\pi$ with ground SBCs from \textsc{sbass}.%
% when the instance size increases, it becomes untractable to parse the program's graph and identify symmetries.
\begin{table}[t]
    \centering
    \rowcolors{2}{gray!25}{white}
    \setlength{\tabcolsep}{7.5pt}
    \resizebox{12cm}{!}{
    \csvloop{
    file=Tables/pigeonHole.csv,
    head to column names,
    before reading=\centering\sisetup{table-number-alignment=center},
    tabular={lrrrrrrr},
    table head=\toprule & \textbf{\textsc{Enum}} & \textbf{\textsc{Sat}} & \textbf{\textsc{Ord}} & \textbf{\textsc{Full}} & \textbf{BASE} &  \textbf{SBASS} & $\mathbf{CLASP^\pi}$\\\midrule,
    command=\Instance & \ABKenum & \ABKsat & \ABKord & \ABKfull & \BASE  &  \SBASS & \Clasp,
    table foot=\bottomrule}}
    \caption{Runtime in seconds for pigeon-hole problem.}
    \label{table:ph}
\end{table}%
\begin{table}[t]
    \centering
    \rowcolors{2}{gray!25}{white}
    \setlength{\tabcolsep}{7.5pt}
    \resizebox{12cm}{!}{
    \csvloop{
    file=Tables/pigeonColor.csv,
    head to column names,
    before reading=\centering\sisetup{table-number-alignment=center},
    tabular={lrrrrrrr},
    table head=\toprule & \textbf{\textsc{Enum}} & \textbf{\textsc{Sat}} & \textbf{\textsc{Ord}} & \textbf{\textsc{Full}} & \textbf{BASE} &  \textbf{SBASS} & $\mathbf{CLASP^\pi}$\\\midrule,
    command=\Instance & \ABKenum & \ABKsat & \ABKord & \ABKfull & \BASE  &  \SBASS & \Clasp,
    table foot=\bottomrule}}
    \caption{Runtime in seconds for pigeon-hole problem with colors.}
    \label{table:pc}
\end{table}%
\begin{table}[t]
    \centering
    \rowcolors{2}{gray!25}{white}
    \setlength{\tabcolsep}{7.5pt}
    \resizebox{12cm}{!}{
    \csvloop{
    file=Tables/pigeonOwner.csv,
    head to column names,
    before reading=\centering\sisetup{table-number-alignment=center},
    tabular={lrrrrrrr},
    table head=\toprule & \textbf{\textsc{Enum}} & \textbf{\textsc{Sat}} & \textbf{\textsc{Ord}} & \textbf{\textsc{Full}} & \textbf{BASE} &  \textbf{SBASS} & $\mathbf{CLASP^\pi}$\\\midrule,
    command=\Instance & \ABKenum & \ABKsat & \ABKord & \ABKfull & \BASE  &  \SBASS & \Clasp,
    table foot=\bottomrule}}
    \caption{Runtime in seconds for pigeon-hole problem with colors and owners.}
    \label{table:po}
\end{table}%

Next, we tested the pigeon-hole problem adding color and owner assignments. 
For the pigeon-hole problem with color assignments, we divided the language bias into two parts:
the first limiting to predicates whose atoms exclusively include variables of the types \texttt{pigeon} and \texttt{hole}, while the second part allows variables to be of the type \texttt{color} too.
Likewise, the problem version with owners and colors required a third language bias extension to variables of the type \texttt{owner}.
For both extensions of the pigeon-hole problem,
the first-order constraints learned in \emph{sat} turned out to be stronger than in the other settings. Nevertheless, all kinds of constraints helped to improve the search for solutions.
Table~\ref{table:pc} and Table~\ref{table:po} show similar results: the constraints learned with the \emph{sat} setting lead to the fastest running times for both satisfiable and unsatisfiable instances. The constraints learned with \emph{enum} using the alternative ordering are shorter and easier to read than the other settings, but sightly less efficient since they break only a subset of all symmetries.
In Table~\ref{table:pc}, the time took for identifying satisfiable and unsatisfiable instances is lower if we use the constraints learned with \textit{fullSBCs} than those learned with \emph{enum}; on the other hand, in Table~\ref{table:po}, we observe the opposite behavior, especially for the last instances:
for the pigeon-hole problem with colors and owners, we could have learned the same constraints in both settings because to obtain the constraints with \emph{enum}, we used instances that identify full SBCs. However, we tested a different set of rules for \textit{fullSBCs} since they were stricter than \emph{enum}, concerning the pigeons' placement symmetries.
Indeed, the first unsatisfiable instance with one color and owner was solved earlier by the constraint of \textit{fullSBCs}.  
Lastly, for small unsatisfiable instances, the ground SBCs from \textsc{sbass} lead to better performance than the constraints learned with the \emph{enum} setting.
However, as soon as the color (or owner) dimension grows, the runs of \textsc{clasp}$^\pi$ reach the timeout.
This behavior is due to the redundancy of the ground SBCs, which slow down the search instead of facilitating it.
For some of the satisfiable instances, finding a solution with the constraints learned in \emph{enum} takes longer than with the original encoding alone. Nevertheless, the latter also has timeouts that do not occur with our learned first-order constraints.%

To conclude, we applied the different settings of our approach to the house-configuration problem \cite{DBLP:conf/confws/FriedrichRFHSS11}, which consists of assigning $t$ things of $p$ persons to $c$ cabinets, where each cabinet has a capacity limit of two things that must belong to the same owner.
Similarly to the pigeon-hole problem with color, we divided the language bias into two parts:
the first limiting to predicates whose atoms exclusively include variables of the types \texttt{cabinet} and \texttt{thing}, while the second part allows variables to be of the type \texttt{person} too.
The running times in Table~\ref{table:hc} exhibit the same trend as observed on the previous problems that our first-order constraints help the search, especially those learned with the \emph{sat} setting. For this problem, the constraints learned with the \emph{enum} setting, the alternative ordering, and exploiting the full SBCs show similar performances.
In some cases, the original encoding is quicker to solve satisfiable instances, although it takes considerably longer for unsatisfiable ones.
On the other hand, \textsc{sbass} brings a moderate speedup for unsatisfiable instances, but its performance suffers a lot when the problem size grows.%
\begin{table}[t]
    \centering
    \rowcolors{2}{gray!25}{white}
    \setlength{\tabcolsep}{7.5pt}
    \resizebox{12cm}{!}{
    \csvloop{
    file=Tables/HouseConfiguration.csv,
    head to column names,
    before reading=\centering\sisetup{table-number-alignment=center},
    tabular={lrrrrrrr},
    table head=\toprule & \textbf{\textsc{Enum}} & \textbf{\textsc{Sat}} & \textbf{\textsc{Ord}} & \textbf{\textsc{Full}} & \textbf{BASE} &  \textbf{SBASS} & $\mathbf{CLASP^\pi}$\\\midrule,
    command=\Instance & \ABKenum & \ABKsat & \ABKord & \ABKfull & \BASE  &  \SBASS & \Clasp,
    table foot=\bottomrule}}
    \caption{Runtime in seconds for house-configuration problem.}
    \label{table:hc}
\end{table}%

%%% Local Variables: 
%%% mode: latex
%%% TeX-master: "ijcai21"
%%% End: 

\section{Conclusions}\label{sec:conclusions}
 This paper introduces methods to lift the SBCs of combinatorial problem encodings in ASP for a target distribution of instances. Our framework addresses the limitations of common \emph{instance-specific} approaches, like \textsc{sbass}, since: 
 \begin{enumerate*}[label=\emph{(\roman*)}]
    \item the knowledge is transferable, as learned constraints preserve the satisfiability 
          for the considered instance distribution; % of the distribution;
      \item the first-order constraints are easier to interpret than ground SBCs;
      \item  the SBCs  are computed offline, allowing for addressing large-\allowbreak{}scale instances, as shown in our experiments; and
      \item the learned  constraints are non-redundant, avoiding performance degradation % of the search 
            due to an excessive ground representation size.
% for problems with more than two dimensions.
     
  \end{enumerate*}
%   We point out that, during the learning phase of
In the current implementation of our approach, \textsc{ilasp} learns shortest constraints that cover as many examples as possible, while there is no distinction regarding the solving performance of candidate hypotheses. Despite this, our experiments showed that the learned constraints significantly improve  the solving performance on the analyzed problems.
%
% previous version sounds like a repetition of the new paragraph in the intro
%
%The two new example generation methods presented in this work implement an alternative atom ordering criterion for the lex-leader scheme and the generation of negative examples for the full symmetry breaking when enumerating all answer sets. Our experiments show that the two novel approaches result in ILP tasks with fewer positive examples and shorter learned constraints, thanks to the \textit{fullSBCs} approach or the alternative atom ordering, respectively.
%
\textcolor{\editcolor}{Moreover, the two example generation methods suggested in this work allowed for ILP tasks with 
\begin{enumerate*}[label=\emph{(\roman*)}]
  \item fewer positive examples and 
  \item shorter learned constraints, 
\end{enumerate*}  
in comparison to the two methods of our previous paper~\cite{tagesc21a}. 
These results are due to the full symmetry breaking when enumerating all answer sets with the \textit{fullSBCs} approach or an alternative atom ordering for the lex-leader scheme, respectively.}

\textcolor{\editcolor}{
Nevertheless, there are still some limitations in the usability of our framework, which partially go back to the components used in our current implementation, i.e., \textsc{sbass}, \textsc{clingo}, and \textsc{ilasp}.
% I would not place speculations here "could be implemented" as it might not be clear for a reviewer how to do so.
The \textsc{sbass} tool does not support ASP programs with weak constraints \cite{cafageiakakrlemarisc19a}, whose implementation is out of the scope of this work.
However, extensions of instance-specific symmetry detection and model-oriented symmetry breaking to optimization problems are undoubtedly worthwhile.
Optimization involves solving unsatisfiable subproblem(s) attempting (and failing) to improve an optimal answer set, where symmetry breaking is particularly crucial for performance. 
%  as that rule type is not supported in its implementation. This issue is out of the scope of the current paper, but it can be addressed with an extension of the tool.
Concerning \textsc{clingo}, if a given encoding $P$ leads to large ground instantiations, the addition of learned constraints does not reduce the size. Therefore, it would be desirable to directly incorporate the information about redundant answer sets into a modified encoding. For instance, for the pigeon-hole problem, this might prevent our method from even generating ground atoms representing the placement of a pigeon into some hole with a greater label.
Lastly, \textsc{ilasp} does currently not scale well with respect to the size of the hypothesis space spanned by the language bias, which is a well-known issue tackled by next-generation ILP systems under development \cite{larubebrlo20a,larubrbe21a}.%
% \todo{Mention conflict analysis?}
    %Conflic analysis in \textsc{ilasp4}: positive example  with empty inclusions and exclusions and many solutions derivable from the context.
    } 

  \textcolor{\editcolor}{
  At present, the successful application of our framework relies on the following characteristics of a combinatorial problem:  \begin{enumerate*}[label=\emph{(\roman*)}]
    \item we can easily provide simple instances (i.e., the total number of solutions can be managed by our implementation) that entail the symmetries of the whole instance distribution; 
    \item the object domains can be expressed in terms of
          unary predicates that hold for a range of consecutive integers; and
    \item the auxiliary predicate definitions suggested for $\mathit{ABK}$ in Section~\ref{subsec:ILPTask} enable the learning of constraints that improve the solving performance.
\end{enumerate*}
In particular, if it gets difficult to compute solutions for an instance in~$S$ to analyze, the formulation of an ILP task to  learn    constraints can become prohibitive.}

 In the future, we aim to investigate whether the learning of SBCs can be readily applied or further adapted to advanced industrial configuration problems, such as the \emph{Partner Units Problem}~\cite{dogalemurisc16a}, as well as complex combinatorial problems with specific instance distributions, like the labeling of \emph{Graceful Graphs}~\cite{petsmi03a}.
 For such application scenarios, the language bias may be enriched, possibly extending the background knowledge with additional predicates characterizing the structure of instances.
 Moreover, for problem instances that yield a vast number of solutions, we can take advantage of the incremental implementation of the \textit{fullSBCs} approach to limit the number of answer sets to consider as examples for an ILP task. 
 Lastly, we intend to develop automatic mechanisms to select suitable instances for $S$ and $\mathit{Gen}$ from instance collections, support lifelong learning, and further optimize the grounding and solving efficiency of learned constraints.

 \paragraph{Acknowledgments.}{This work was partially funded by
 KWF project 28472,
 cms electronics GmbH,
 FunderMax GmbH,
 Hirsch Armbänder GmbH,
 incubed IT GmbH,
 Infineon Technologies Austria AG,
 Isovolta AG,
 Kostwein Holding GmbH, and
 Privatstiftung Kärntner Sparkasse.
 We thank the anonymous reviewers for their helpful and constructive comments.}
%%% Local Variables: 
%%% mode: latex
%%% TeX-master: "ijcai21"
%%% End: 

%\begin{acknowledgements}
%If you'd like to thank anyone, place your comments here
%and remove the percent signs.
%\end{acknowledgements}

% Authors must disclose all relationships or interests that 
% could have direct or potential influence or impart bias on 
% the work: 
%
% \section*{Conflict of interest}
%
% The authors declare that they have no conflict of interest.

% BibTeX users please use one of
%\bibliographystyle{spbasic}      % basic style, author-year citations
%%% \bibliographystyle{spmpsci}      % mathematics and physical sciences
%\bibliographystyle{spphys}       % APS-like style for physics
%%% \bibliography{include/bibliography/krr,mybib,include/bibliography/procs}  % name your BibTeX data base

% Non-BibTeX users please use

\end{document}